\documentclass{bmcart}

\usepackage{braket}
\usepackage{amsmath,amsfonts,amssymb,amsthm,pifont}
\usepackage{amsbsy,amscd,mathrsfs,graphicx,color}
\usepackage{color}
\RequirePackage[numbers]{natbib}
\RequirePackage{hyperref}
\usepackage[utf8]{inputenc} 

\hypersetup{%
  pdfpagemode=UseNone, 
  pdfstartpage=1,
  pdfstartview=FitH,
  pdftoolbar=true,
  colorlinks = true,
  linkcolor=blue,
  citecolor=blue,
  bookmarksopen=false
}

\startlocaldefs

\newcommand{\openone}{\leavevmode\hbox{\small1\kern-3.3pt\normalsize1}}

\definecolor{darkorange}{rgb}{1,0.5,0.14} 
\endlocaldefs

\begin{document}

\begin{frontmatter}

\begin{fmbox}
\dochead{Review}


\title{Quantum optimal control in quantum technologies. Strategic report on current status, visions and goals for research in Europe}


\author[
  addressref={aff1},                   
  corref={aff1},                       
  email={christiane.koch@fu-berlin.de}   
]{\inits{C.P.}\fnm{Christiane P.} \snm{Koch}}
\author[
  addressref={aff2},
  email={ugo.boscain@cmap.polytechnique.fr}
]{\inits{U.}\fnm{Ugo} \snm{Boscain}}
\author[
  addressref={aff3},
  email={tommaso.calarco@fz-juelich.de}
]{\inits{T.}\fnm{Tommaso} \snm{Calarco}}
\author[
  addressref={aff4},
  email={dirr@mathematik.uni-wuerzburg.de}
  ]{\inits{S.}\fnm{Gunther} \snm{Dirr}}
\author[
  addressref={aff5,aff5a,aff7},
  email={Stefan.Filipp@wmi.badw.de}
  ]{\inits{S.}\fnm{Stefan} \snm{Filipp}}
\author[
  addressref={aff6,aff7},
  email={glaser@tum.de}
  ]{\inits{S.J.}\fnm{Steffen J.} \snm{Glaser}}
\author[
  addressref={aff11},
  email={ronnie@fh.huji.ac.il}
  ]{\inits{R.}\fnm{Ronnie} \snm{Kosloff}}
\author[
  addressref={aff9},
  email={simone.montangero@unipd.it}
]{\inits{S.}\fnm{Simone} \snm{Montangero}}
\author[
  addressref={aff6,aff7},
  email={tosh@tum.de}
  ]{\inits{T.}\fnm{Thomas} \snm{Schulte-Herbr\"uggen}}
\author[
  addressref={aff8},
  email={Dominique.Sugny@u-bourgogne.fr}
  ]{\inits{D.}\fnm{Dominique} \snm{Sugny}}  
\author[
  addressref={aff3,aff10},
  email={f.wilhelm-mauch@fz-juelich.de}
]{\inits{F.K.}\fnm{Frank K.} \snm{Wilhelm}}


\address[id=aff1]{
  \orgdiv{Dahlem Center for Complex Quantum Systems and Fachbereich Physik}, 
  \orgname{Freie Universit\"at Berlin},          
  \street{Arnimallee 14},
  \postcode{14195}
  \city{Berlin},                              
  \cny{Germany}                                    
}
\address[id=aff2]{
  \orgdiv{Laboratoire Jacques-Louis Lions}, 
  \orgname{Sorbonne Universit\'e, Universit\'e de Paris, CNRS, Inria},   
  \city{Paris},                              
  \cny{France}                                    
}
\address[id=aff3]{%
  \orgdiv{Peter Gr\"unberg Institut},
  \orgname{Forschungszentrum J\"ulich},
  \city{J\"ulich},
  \cny{Germany}
}
\address[id=aff4]{
  \orgdiv{Institut f\"ur Mathematik}, 
  \orgname{Universit\"at W\"urzburg},          
  \city{W\"urzburg},                              
  \cny{Germany}   
}
\address[id=aff5]{
  \orgdiv{Walter-Meissner-Institut}, 
  \orgname{Bayerische Akademie der Wissenschaften},   
  \city{Garching},                              
  \cny{Germany}                                    
}
\address[id=aff5a]{
  \orgdiv{Department Physik}, 
  \orgname{Technische Universit{\"a}t M{\"u}nchen (TUM)},   
  \city{Garching},                              
  \cny{Germany}                                    
}
\address[id=aff6]{
\orgdiv{Department Chemie},
  \orgname{Technische Universit{\"a}t M{\"u}nchen (TUM)},   
  \city{Garching},                              
  \cny{Germany}                                    
}
\address[id=aff7]{
  \orgname{Munich Centre for Quantum Science and Technology (MCQST)}, 
  \city{M\"unchen},                              
  \cny{Germany}                                    
}
\address[id=aff11]{
  \orgdiv{The Institute of Chemistry}, 
  \orgname{The Hebrew University of Jerusalem},          
  \city{Jerusalem},                              
  \cny{Israel}  
}     
\address[id=aff9]{
  \orgdiv{Department of Physics and Astronomy 'G. Galilei'}, 
  \orgname{University of Padova},   
  \city{Padova},                              
  \cny{Italy}                                    
}
\address[id=aff8]{
  \orgdiv{Laboratoire Interdisciplinaire Carnot de Bourgogne (ICB)}, 
  \orgname{Universit\'e de Bourgogne-Franche Comt\'e},   
  \city{Dijon},                              
  \cny{France}                                    
}
\address[id=aff10]{
  \orgdiv{Theoretical Physics}, 
  \orgname{Universit\"at des Saarlandes},          
  \city{Saarbr\"ucken},                              
  \cny{Germany}  
}                                  



\end{fmbox}


\date=\today
\begin{abstractbox}

\begin{abstract} 
%
Quantum optimal control, a toolbox for devising and implementing the shapes of external fields that accomplish given tasks in the operation of a quantum device in the best way possible, has evolved into one of the cornerstones for enabling quantum technologies. The last few years have seen a rapid evolution and expansion of the field. We review here recent progress in 
our understanding of the controllability of open quantum systems and in the development and application of quantum control techniques to quantum technologies. We also address key challenges and sketch a roadmap for future developments.
\end{abstract}




\end{abstractbox}
%

\end{frontmatter}



\section{Introduction}\label{sec:intro}

Quantum optimal control theory (QOCT) refers to a set of methods to devise and implement shapes of external electromagnetic fields that manipulate quantum dynamical processes at the atomic or molecular scale in the best way possible~\cite{glaserEPJD2015}. It builds on control theory in more general terms which evolves at the interface between applied mathematics, engineering, and physics and concerns the manipulation of dynamical processes to realize specific tasks. The main goal is for the dynamical system under study to operate optimally and reach its physical limits while satisfying constraints imposed by the devices at hand. Quantum processes are no exception to this general framework, but certain aspects of control theory must be adapted to take into account the particularities of the quantum world. Over the past few years, QOCT has become an integral part of the emerging quantum technologies~\cite{Acin2018}, testifying to the fact that it is control that turns scientific knowledge into technology~\cite{glaserEPJD2015}:
If the superposition principle is the core feature of quantum mechanics, quantum control is the superposition principle at work. 

Quantum technologies require comparatively well-isolated and well-characterized quantum systems. It is this very feature that makes them an ideal testbed for QOCT, compared to other fields where QOCT has been used, such as chemical reaction dynamics. Conversely, QOCT has matured to the stage that it is nowadays readily used in experiments. The next challenge for QOCT will be to become an integral part of practical quantum devices or, in other words, of practical quantum engineering. 

Here we provide an update to
Ref.~\cite{glaserEPJD2015}, focussing on progress in QOCT and its applications relevant to the development of quantum technologies. Under this specific perspective, quantum optimal 
control sets out to answer typical engineering questions. For example: To which extent can a quantum system be (i) controlled, (ii) observed (sensed or tomographed), (iii) stabilised, etc. For classical (mostly linear) systems, a rigorous systems and control theoretical framework exists and is core to the teaching programme of every engineer. For training future quantum engineers, such a framework is yet to emerge. 

A rigorous and unified quantum systems theory is therefore among the current overarching research goals --- it will interface not only theory and experiment but teaching programmes in quantum physics and engineering as well. Such a theory also forms the basis for the derivation of optimal control strategies by ensuring the well-posedness of problems and existence of solutions. We provide a brief summary of basic definitions together with a review of recent progress towards these goals and open questions in Sec.~\ref{sec:controllability}. 

Section~\ref{sec:methods} presents the current state of the art in quantum optimal control \textit{methodologies}. These can be classified into analytical vs numerical approaches and the latter into approaches evaluating only the target functional (gradient-free methods) and those based on variational calculus such as the Pontryagin maximum principle (PMP). We review progress on these methodologies in Sec.~\ref{sec:methods}, including corresponding software development, for which we highlight publicly available software packages.

Quantum optimal control is closely related, both in terms of goals and techniques, to several neighbouring fields, including most notably quantum feedback control, machine learning, and shortcuts to adiabaticity. We highlight similarities and differences between quantum optimal control theory and these approaches in Sec.~\ref{sec:comparison}, pointing also to recent fruitful cross-fertilization. An example of this is the inclusion of ideas from both quantum feedback and machine learning to quantum optimal control, in order to account for model inaccuracies and enhance practical applicability of the approach.

\begin{figure}\label{fig:QCBEC}
\includegraphics[scale=0.38]{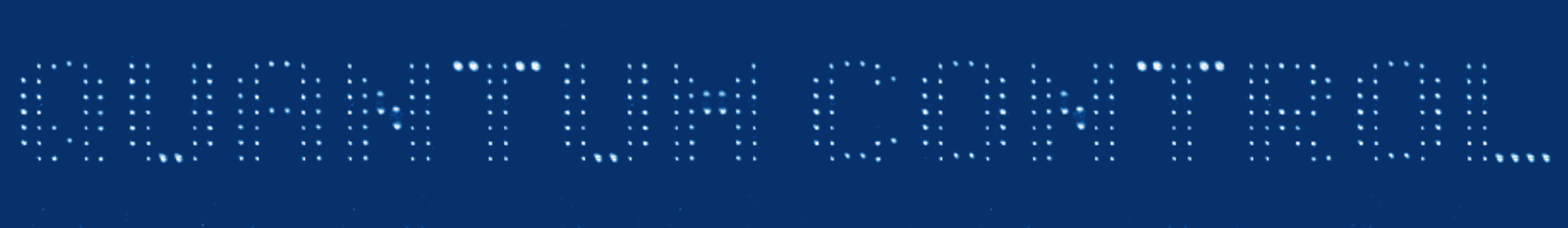}
\caption{A BEC "printer": The robust and versatile control of a BEC allows for producing experimentally lines of points made up of atomic clouds, with which letters and words can be formed.}
\end{figure}

When looking back to the start of the art presented in Ref.~\cite{glaserEPJD2015}, scientific progress has been most impressive in the number and extent of practical quantum technological applications exploiting quantum optimal control. We review these advances in Sec.~\ref{sec:applications}, starting with experimental demonstrations of quantum optimal control. A recent and striking example of the power of quantum optimal control techniques is illustrated in Fig.~\ref{fig:QCBEC} showing the realization of a Bose-Einstein condensate (BEC) printer~\cite{dupont2021}. In this experiment, a BEC of ultra-cold $^{87}$Rb atoms was loaded into a one-dimensional optical lattice formed by two counter-propagating laser beams with the same wavelength, but a different phase. A gradient-based optimal control algorithm was used to calculate the phase which optimally "shakes" the optical lattice back and forth and thus brings the quantum system to the desired target state. At the end of the control process, the atoms are in a well-defined superposition of speeds, which are multiples of an elementary speed. This superposition can be experimentally visualized through a ballistic expansion achieved after switching off the confining potential. The measured chain of small atomic clouds allows one to write line by line, for example \emph{Quantum Control} as shown in Fig.~\ref{fig:QCBEC}. This approach to preparing states of a BEC in an optical lattice is also practically useful in many areas from quantum simulation to quantum metrology~\cite{dupont2021}.
We review further highlights of experimental implementations of quantum optimal control in Sec.~\ref{sec:applications} and then provide an overview over quantum optimal control approaches tailored to, respectively, specific quantum hardware and specific key tasks in the operation of quantum devices. This section also elucidates the relation between quantum control and quantum thermodynamics. A summary of goals and challenges in view of further expanding the scope of quantum optimal control applications in practical quantum devices in the mid-term completes Sec.~\ref{sec:applications}.

Finally, we discuss the goals and challenges that define our vision for the longer term development of quantum optimal control in the quantum technologies in Sec.~\ref{sec:vision} and conclude in Sec.~\ref{sec:conclusions}.

\section{Controllability and accessibility of open quantum systems}\label{sec:controllability}

Natural and foremost questions for engineering quantum technological devices 
are `what can one do with them?',  in particular `which states can be prepared?' or `which quantum gates can be implemented?' Answering these questions connects engineering with the core of mathematical control theory.
This section starts by giving an account on how to formalise these questions mathematically, adapting the classical engineering terminology of controllability, accessibility and reachable sets to the realm of
closed and open quantum systems. To this end, we start out with the necessary basic definitions in Sec.~\ref{sec:Theory-Defs} to describe the
pertinent recent progress
for controlling symmetric systems in Sec.~\ref{sec:symmetric-systems} and
for reachable sets in Markovian systems in Sec.~\ref{sec:Reach-Markov-Open}. 
In a wider context, we characterise the role of Markovianity for control in Sec.~\ref{sec:Markovianity}.
The loss of compactness in (finite dimensional) open systems paves the way to the even more intricate case
of infinite dimensional closed systems treated in Sec.~\ref{sec:infinite-dim-H}.
The problem of simultaneous robust control of infinitely many (almost identical) subsystems in Sec.~\ref{sec:Sim-Control} 
may be seen in close spirit.
Time-optimal control problems are put into the context of the quantum-speed limit in Sec.~\ref{sec:Q-Speed-Limit}.
Finally, the progress is wrapped up to serve as a roadmap to challenging
goals and open research problems formulated in Sec.~\ref{sec:Theory-Goals}. 

\subsection{Basic definitions}\label{sec:Theory-Defs}

A control system is usually described by an equation of motion,  e.g. an ordinary or partial differential equation, involving additional ``parameters'' ({\em controls}). The controls can be time-dependently
manipulated to ``steer'' the system, say from a given initial state to a desired target state.  The system is called {\em controllable} if any initial state can be transformed into any desired target state. The {\em reachable set} of a given initial state is composed of all states the initial state can be steered to. In other words, a system is controllable if  the reachable set to any initial state coincides with the entire state space. The system is {\em accessible} if its reachable set contains at least interior points. Roughly speaking, for an accessible system, the reachable sets may be small
but not too small in the sense that they embrace at least a set of full dimension. In the following, we characterize  standard quantum control scenarios and their properties in terms of the notions introduced above.

A typical closed quantum control system extends the Schr\"odinger equation (just governed via its so-called drift Hamiltonian) 
by several control Hamiltonians whose impact is scaled by possibly time-dependent control amplitudes---for instance,
$u_x(t)\sigma_x$ for an $x$-pulse of amplitude $u_x(t)$ on a single qubit.
With the controlled part 
being linear in {\em both} the state and the control, it is an example of the  wide class of {\em bilinear control systems}~\cite{Elliott09,Jurdjevic97,dAlessandroBook2008}. Before investigating controllability, one has to keep in mind that one can associate different state spaces to a quantum control problem: pure states, mixed states,  and unitary gates.  An easy and well-established criterion whether a finite dimensional quantum dynamical system of such bilinear form is fully (i.e., unitary gate) controllable proceeds via the {\em system Lie algebra} obtained as linear span generated by all iterated commutators among system and control Hamiltonians (multiplied by $i$): if it amounts to the full Lie algebra $su(N)$, generating the special unitary group $SU(N)$ by exponentiation, then the system is fully controllable (also called {\em universal}). This widely used criterion is termed `Lie-algebra rank condition' ({\sc larc}) \cite{JS72,Bro73,Jurdjevic97}. It exploits the fact that (due to compactness of $SU(N)$) the trajectory generated by the drift Hamiltonian is (almost) periodic, and therefore forward and backward time evolution of the drift can be used to steer the system. When restricting to pure states in even dimensions~$N$, it suffices that the system Lie algebra yields the Lie algebra of all unitary symplectic matrices of dimension $N$. Otherwise, in finite dimensional closed systems, pure-state controllability, mixed-state controllability, and full unitary gate controllability coincide~\cite{AlbertiniTAC2003,DirrGAMM2008,KDH12}. In infinite dimensions, the situation is more involved and in spite of recent progress elucidated in Sec.~\ref{sec:infinite-dim-H} and Sec.~\ref{sec:Sim-Control}, a controllability condition  as powerful as {\sc larc} is still sought for, see Sec.~\ref{sec:Theory-Goals}.

Already {\em open quantum systems} in finite dimensions are more complicated. Here we focus on the Markovian scenario, i.e., on master equations taking the standard form of a controlled Gorini-Kossakowski-Sudarshan-Lindblad equation \cite{Koss72,Davies74,GKS76,Lind76,DirrRMP2009}~(GKLS).
A {\em caveat} in advance:
If the GKLS reduced dynamical equation is derived under the additional assumption that the total of system plus environment 
in the sense of Stinespring has time-translation symmetry
(i.e. the total unitary commutes with the sum of system and bath Hamiltonian), then one arrives at {\em enhanced thermal Markovian operations}~\cite{Lostaglio19r,CwiklinskiHorodeckiOppenheim15}. 
Alternatively, this is implied by Markovian evolutions respecting {\em strict energy conservation} as shown in~\cite{dann2021open},
where {\em thermodynamic compatibility} imposes a functional dependence between the dissipative and unitary generators.
It should be emphasized that in general, the GKLS master
equation {\em per se} need not obey thermodynamical principles \cite{levy2014local,pekola2021}, cf.~Sec.~\ref{subsec:quthermo}. An example of physical noise that does {\em not} meet thermodynamic compatibility is standard bit-flip, while standard phase-damping does.

Possible state spaces for controllability analysis in open systems are the set of all density operators (as the irreversible time evolution of open quantum systems no longer preserves the spectrum of the initial density operator), or the set of all {\em quantum maps},  i.e., the set of all completely positive and trace-preserving ({\sc cptp}) operators. These maps ensure density operators to evolve into density operators; they form a {\em semi}\/group --- not a group as
unitary propagators do in closed systems. This issue will be further discussed below in Sec.~\ref{sec:Markovianity}. Here we just note that the reachable set of open Markovian dynamics takes the form of a {\em Lie-semigroup} orbit~\cite{Lawson99,DirrRMP2009} generated by the associated Lie wedge~\cite{HHL89}, whereas in closed systems the reachable set generically takes the form of a Lie-group orbit generated by the
associated system Lie algebra. Remarkably, the set of all (time-dependent) {\em Markovian} quantum maps carries the structure of a Lie semigroup \cite{DirrRMP2009,OSID17}, whereas the entire set of all quantum maps (with positive determinant) also embracing non-Markovian ones does not. 

By irreversibility, open systems with permanent noise are not exactly (mixed-state) controllable~\cite{Alt04,DirrGAMM2008}. However, generic finite dimensional open quantum systems with (usually Hamiltonian) controls accompanying a (usually non-switchable) relaxation term are accessible. The concept of accessibility, which is considerably weaker than controllability, is nevertheless a good starting point for characterising reachable sets in open quantum dynamics. Any finite-dimensional Markovian open quantum system has at least one fixed point, the steady state under the drift Hamiltonian plus dissipator. If the identity matrix $\openone_N$ and, by linearity, the maximally mixed state $\frac{1}{N}\openone_N$, is among the fixed points, the map is termed
{\em unital}. A generic finite-dimensional unital system is accessible~\cite{Alt04, DirrGAMM2008, KDH12},
if its system Lie algebra 
is isomorphic to the Lie algebra $\mathfrak{gl}_{N^2-1}(\mathbb{R})$ of all  real square matrices with $N^2-1$ rows and columns
(details of the non-unital case are discussed in  Refs.~\cite{DirrGAMM2008, KDH12}). This criterion is
likewise called Lie-algebra rank condition ({\sc larc}), the difference between closed and open systems being that in closed systems {\sc larc} is equivalent to controllability,  while in open systems it is only equivalent to accessibility. This contrast results from different compactness properties of the underlying Lie groups $SU(N)$ and $GL_{N^2-1}(\mathbb{R})$: While the compactness of $SU(N)$ forces the one-parameter semigroup generated by the
drift term to be (almost) periodic and thus time-reverting, this argument fails for the dissipative term in open systems.

\subsection{Recent progress on the controllability of open quantum systems}

\subsubsection{Controllability within symmetry-induced subsystems}\label{sec:symmetric-systems}
Moving from controllability with respect to the entire state space to controllability with respect to a symmetry-adapted state space has recently been exploited for spin systems with permutation symmetry~\cite{albertini2021,AA21,AH21}. 
More precisely, by means of a Clebsch-Gordan decomposition, one arrives at block diagonal Hamiltonians in a symmetry-adapted basis, where controllability can naturally be discussed as subspace controllability for every block~\cite{AA18} (see also~\cite{SS09}). 
Loosely connected in spirit is another recent development~\cite{vanvriedvelde2021} which circumvents implementing controlled unitary gates by resorting to Kraus maps implementing unitary gates or channels just on a $d$-dimensional subset of qubits, while leaving the remaining qubits invariant. These channels are thus called sector preserving and they promise to widen the set of implementable quantum circuits.

\subsubsection{Reachable sets for open Markovian systems with switchable noise}\label{sec:Reach-Markov-Open}
As mentioned before, open systems with unitary control and {\em non-switchable} noise are never controllable. On the other hand, under the assumptions that (i) the respective noise term can be turned on and off and (ii) the ``residual'' closed system (with the dissipator switched off) is fully unitarily controllable, one can precisely characterize reachable sets for certain classes of {\em unital} and {\em non-unital} systems:
In {\em unital systems} with full unitary control and a single 
bang-bang switchable noise generator, the reachable set to any initial density operator is given by all density operators majorized~\footnote{In a pair of Hermitian matrices,  $A$ majorizes $B$ if all partial sums over the eigenvalues of $A$ sorted by descending magnitude are larger or equal than those of $B$.} by the initial one~\cite{vomEndeOSID2019} 
(details in~\cite{vEnde-PhD2020}).
Here, coherent and
incoherent controls are allowed to operate on different time scales, so the result formally reproduces earlier findings with instant unitary controls~\cite{Yuan10}.

Recently, first generalisations of these results to {\em infinite dimensional} systems with a single switchable and compact noise generator~\cite{vomEndeOSID2019} were obtained. Yet, a better understanding of infinite dimensional systems incorporating earlier work on open systems with unbounded drift or noise~\cite{Somaraju_Mirrahimi_Rouchon13} requires further work.

For {\em non-unital systems} with a switchable Markovian coupling to a bath at temperature $T=0$, one obtains controllability on the set of all density operators~\cite{dirr2019}. A switchable cooling to low temperatures 
can readily be implemented experimentally, for instance by a tunable coupling to an open transmission-line in the specific GMon setting of~\cite{Mart14},
where the local cooling part itself can be made to respect conditions for Markovianity 
as well as for enhanced thermal operations~\cite{BSH16}.
First generalisations of cooling via baths of temperatures $0<T<\infty$ (again in the realm of enhanced thermal operations) have
recently been obtained for a toy model restricting the dynamics to population transfer in the (cooling-preserved) eigenbasis of the drift Hamiltonian~\cite{dirr2019} thus respecting strict energy conservation. Using generalized majorisation techniques~\cite{vomEnde19polytope}, the reachable sets of the restricted model could be upper estimated~\cite{MTNS20}. Similar ideas discussed in earlier studies~\cite{rooney2018} are considerably more difficult to handle explicitly in the general case, whereas two-level systems can be treated exactly~\cite{rooney2012} even allowing for
giving explicit reachable sets~\cite{lokutsievskiyJPA2021}. 
Restricted to single two-level systems, the latter describes an intermediate scenario with coherent and incoherent controls, where full decoupling of system and bath cannot be achieved.

\subsubsection{Markovianity and its role in quantum systems and control theory}\label{sec:Markovianity}
In terms of state transfer, non-Markovian control systems can be mimicked by Markovian ones with switchable coupling to a bath at $T=0$~\cite{BSH16,MTNS20} so that their reachable sets can essentially coincide. Yet non-Markovian transfer may at instances be more efficient~\cite{HuelgaPlenio21}. On the operator level, however, this coincidence 
no longer holds: there are quantum maps in the Kraus representation that cannot be represented as solutions of Markovian master equations of {\sc gkls}-form --- these are non-Markovian (i.e.\ neither time-dependent nor time-independent Markovian).
Progress has been made in characterising non-Markovianity~\cite{Plenio_NMarkov_Review2014} in particular by analysing information backflow and structured environmental spectral densities~\cite{Breuer_NMarkov_Review2016}, and a hierarchy 
for abundant definitions of (non)Markovianity has been set up~\cite{Wiseman_Markov2018}.
In contrast to standard Markovian master equations (of {\sc gksl} form), 
non-Markovian master equations come with different types of memory kernels~\cite{DiosiFerialdi_NMarkovMasterEqn2014,Ferialdi_NMarkovMasterEqn2016,Vacc_NMarkovMasterEqn2016,ChruscinskiKossakowksi2016,Chruscinski2019}.
Non-Markovian control systems thus depart from the bilinear setting with its
clear correspondence between generators and propagators of time evolution
exploited to assess controllability (respectively accessibility) on the generator level, which made the Markovian case~\cite{Lawson99,DirrRMP2009} discussed above so convenient.
Moving to the level of Kraus maps instead is more involved.
To our knowledge, the only explorations of controllability at the level of Kraus maps~\cite{Rabitz07b, Pechen11} have been performed without comparing the non-Markovian reachable sets to their restrictions under Markovian conditions.

\subsubsection{Controllability of closed quantum systems evolving on infinite dimensional Hilbert spaces}\label{sec:infinite-dim-H}

The loss of compactness mentioned in Sec.~\ref{sec:Reach-Markov-Open} already 
occurs in {\em closed} infinite dimensional systems~\cite{keyl18InfLie, vomEndeOSID2019,duca2021bilinear,Duca2020}. As a result, if the underlying Hilbert space is infinite dimensional, one can only expect approximate controllability. This refers to controllability in the sense that one can reach every state arbitrarily closely but not necessarily exactly~\cite{ball1982controllability,CaponigroChambrion20,boscain2015approximate}. The common assumption that the Hamiltonian drift term has a discrete spectrum suggests to use finite dimensional (Galerkin-type) approximations~\cite{boscain2015approximate,boscain2015control}. This allows to employ recurrence arguments similar to the finite-dimensional case~\cite{keyl18InfLie,chambrion2009controllability,boscain2012weak,boscain2014multi}. If the spectrum of the drift is non-resonant, i.e., all energy gaps are different, controllability analysis is comparatively straightforward~\cite{boscain2012weak}. In contrast, many degeneracies make the controllability analysis harder~\cite{boscain2014multi}. Based on~\cite{boscain2014multi}, the conditions for completely controlling the rotational degrees of freedom of molecules, a quantum technology platform of renewed interest~\cite{albertPRX2020}, have been identified~\cite{boscainJCO2021,leibscher2020,pozzoli2021}. 

Systems with continuous spectrum---no matter whether closed or open---are even more delicate. The typical scenario is that of a continuous variable without  confining potential and the corresponding controllability analysis has long been considered as particularly difficult~\cite{chambrion2012periodic,boussaid2015approximate}. A drift Hamiltonian with continuous spectrum also arises from a parameter-dependent Schr\"odinger equation, where the parameters describe some model uncertainties/inhomogeneities. This scenario, referred to as {\em simultaneous or ensemble control}, is discussed  in the next subsection.

\subsubsection{Simultaneous controllability}\label{sec:Sim-Control}
Control of an ensemble of quantum systems is of particular interest for quantum engineering because it allows to gain robustness of control procedures without feedback techniques. Starting with the seminal paper~\cite{likhaneja2009}, ensemble control has become an active research field in control theory. In the simplest case of finite parameter sets, the ensemble-control problem reduces to simultaneously controlling a finite number of almost identical systems. If the individual subsystems evolve on $SU(N)$, simultaneous controllability was first investigated in \cite{Altafini09} and fully characterized in terms of Lie algebraic conditions~\cite{dirr2012,Turinici15}. Controllability in the case that the parameters can assume infinitely many different values, either in a countable set \cite{Chittaro18} or in a non-trivial compact set, is much harder.
The latter setting with its important applications in robust quantum  control~\cite{KhanejaGlaser2006,ansel2020} has 
led to a novel branch in non-linear control theory \cite{Agrachev16,Chen19}.
Roughly, one can distinguish two different approaches: (i) infinite dimensional Lie group techniques and (ii) adiabatic methods~\cite{boscain2015approximate}.
Ensemble controllability has been studied for the infinite-dimensional case \cite{Zhang2021} and efficient numerical algorithms have been developed both for fixed-endpoint and free-endpoint control problems \cite{Wang2017,Wang2018}.

(i) For casting the simultaneous control problem into the setting of infinite dimensional Lie groups, one chooses as state space all square integrable functions over the compact parameter range $K \subset \mathbb{R}^m$ with values in the Hilbert space common to all subsystems, e.g., $L^2(K,\mathbb{C}^N)$. 
The corresponding unitary group contains the infinite dimensional Lie subgroup which consists of all continuous maps from $K$ to $SU(N)$.
This subgroup takes to role of $SU(N)$ in ensemble control. Ref.~\cite{likhaneja2009} showed that one can achieve approximate controllability (with respect to above Lie subgroup) for a very particular set of parameter-dependent generators of $SU(2)$.
A similar result was obtained for the Bloch equation~\cite{beauchardCMP2010}  
while \cite{Agrachev16} generalized these ideas to the class of control-affine systems including bilinear systems. 
Later \cite{Chen19} proved that every semi-simple Lie algebra allows for a special set of parameter-dependent generators such that the ideas of \cite{likhaneja2009} can be carried over. This is important from an engineering perspective
because it says that one can design the control Hamiltonians of a finite dimensional system such that approximate simultaneous controllability can be achieved---yet the number of necessary control Hamiltonians is in general quite large. 
However, it is still an open problem to what extent one of the control fields can be replaced by the drift term without loosing controllability, see Sec.~\ref{sec:Theory-Goals}.

(ii) Adiabatic control is well known for its properties of robustness against dispersion of system parameters. Typical pulse designs based on these ideas are chirped pulses~\cite{Shore1990} and counter-intuitive pulses for
{\sc stirap} processes~\cite{Bergmann2021}. Many more different control protocols based on adiabatic passage were proposed in the last three decades by physicists~\cite{guerinACP2003}. The mathematical analysis developed in \cite{augierJCO2018} permits to understand and prove rigorously that, when acting with two controls, robustness results from the presence of conical intersections between energy levels. 
Such eigenvalue intersections spread into a curve in presence of a dispersion parameter that one can ``follow''  adiabatically and thus obtain population transfer for every value of the parameter. The price to pay is an adiabatic transfer at the order $\sqrt{\varepsilon}$ instead of $\varepsilon$
meaning that in a time of order $\varepsilon^{-1}$ one obtains a transfer up to errors of order  $\sqrt{\varepsilon}$ instead of $\varepsilon$. 
An extension of the adiabatic protocols based on the inertial theorem  \cite{dann2021inertial} shares the property of 
\/``time-dependent\/'' constants of motion. 
The robustness property of the protocol has been studied experimentally \cite{hu2021experimental}.
Using this idea, simultaneous controllability can be realized for a large class of systems. However, how to extend these ideas to obtain simultaneous operator controllability is an open question.

In many situations, more than one control is necessary for simultaneous controllability~\cite{Agrachev16,beauchardCMP2010,likhaneja2009}. For example, a typical scenario involves two controls that are obtained after a rotating-wave approximation. However, compatibility of the rotating-wave approximation with adiabatic theory is a problem overlooked for a long time (see the discussion  in Ref.~\cite{rouchon2008quantum}). The compatibility of the two approximations has recently been studied in detail~\cite{augier2020,robin2020compatibility}.
In particular for a qubit driven by a single control, the range of dispersion of the Larmor frequency allowing for simultaneous controllability was identified. These ideas open new perspectives for simultaneous controllability under a single control field. While all these findings show that simultaneous control can be theoretically accomplished, reliable error estimates for numerical investigations, which are often based on standard algorithms applied to a finite parameter sample, are still missing.

\subsubsection{Quantum speed limit}\label{sec:Q-Speed-Limit}
%

The presence of a drift Hamiltonian that does not belong to the Lie algebra generated by the control Hamiltonians implies the existence of a system intrinsic timescale. As a result, arbitrarily strong fields are not sufficient to speed up the system dynamics, and it is impossible to prepare a desired state or carry out a desired quantum gate in arbitrarily short time \cite{Khaneja2001, Khaneja2002}. The duration of a {\it "time-optimal control sequence" } is called the minimum time for the control task \cite{Khaneja2001} and  is sometimes also referred to as controllability time \cite{AgrCha06} or
the  {\em quantum speed limit} ({\sc qsl})~\cite{deffnerJPA2017,poggiAFA2020}. 
A vanishing controllability time or, equivalently, a  diverging quantum speed produces a reduced uncertainty in quantum observables, and it can be understood as a consequence of emerging classicality for these particular observables~\cite{poggiPRXQ2021}. Beyond two-level and three-level systems, the {\sc qsl} is most often determined numerically in a heuristic way, by lowering the control time~\cite{canevaPRL2009, KHANEJAJMR2005, Schulte2005} 
Recently, an alternative approach has been introduced where the {\sc qsl} is determined by transforming the quantum control problem to a quadratically constrained quadratic program with generalized probability conservation laws as the constraints and relaxation of the quadratic to a semidefinite program~\cite{zhangPRL2021}. Yet another alternative, applicable to Hermitian and non-Hermitian quantum systems, determines the  {\sc qsl} by the changing rate of phase which represents the transmission mode of the quantum states over their evolution~\cite{sunPRL2021}. 
A {\sc qsl} for relative entropies between the output and the input has been derived for a general unitary channel~\cite{piresPRE2021}, and the {\sc qsl} for pure state entanglement corresponds, not too surprisingly, to the minimal time necessary to unitarily evolve a given quantum state to a separable one~\cite{rudnickiPRA2021}. For finite dimensional systems, the bound can also be expressed in terms of rotations on the Bloch sphere~\cite{liuPRA2021}.
Analysis of the {\sc qsl} can be extended to a quantum mechanical treatment of the external control~\cite{gherardini2020}. The {\sc qsl} has been suggested as a measure of robustness~\cite{kobayashiPRA2020} and as means to characterize the reachable set of states~\cite{arenzNJP2017,kobayashi2021}.

For open quantum systems, the {\sc qsl} is most often determined by the dissipative timescales~\cite{uzdinEPL2016}.
More precisely, under the assumption that states with the same purity
can be reached in arbitrarily short time, a speed limit can be derived which only depends on the relaxation rates~\cite{uzdinEPL2016,diazPRA2020}. Mixed states are relevant also when estimating the speed limit bound in the classical limit~\cite{bolonekQuantum2021}. While  for specific classes of dynamical evolutions and initial states, a link between non-Markovianity of the dynamics and the {\sc qsl} exists, this is not true in general~\cite{teittinenNJP2019,teittinenEntropy2021}. The {\sc qsl} has been connected to thermodynamic quantities such as energy fluctuations or the entropy production rate for Markovian and  non-Markovian dynamics~\cite{funoNJP2019,dasPRA2021}. The usual geometric approach interpreting the {\sc qsl} as a consequence of the metric on the state space can be complemented by action quantum speed limits~\cite{oConnorPRA2021}. These depend on the instantaneous speed with which a path in state space is traversed and can also be used  as an indicator for optimality of the path~\cite{oConnorPRA2021}.

Another challenge is the identification of the {\sc qsl} in many-body systems:
When only local controls are allowed, the controllability time can be exponentially large in the system size due to a diverging geodesic length~\cite{bukovPRX2019}. This is related to an extensively growing sensitivity of the many-body system to local perturbations which in turn can be fully characterized by the {\sc qsl}~\cite{fogartyPRL2020}. For systems undergoing quantum phase transitions, the {\sc qsl} is obtained for 
counterdiabatic driving which turns out to encode the Kibble-Zurek mechanism~\cite{pueblaPRR2021}. Mixed states of many-body systems require particular care since the  {\sc qsl}, typically, is dramatically overestimated but for thermal states in a closed many-body system a meaningful bound can be derived in the thermodynamic limit~\cite{ilinPRA2021}. 

The results concerning the quantum speed limit described above concern systems evolving on finite dimensional Hilbert spaces. For systems evolving on infinite dimensional Hilbert spaces the problem is more subtle.
 For example, there exist systems for which the drift Hamiltonian is not generated by the controlled ones, that have a controllability time that can be reduced to zero~\cite{boussaidIEEECDC2012}. In contrast for a charged particle in a magnetic field, the controllability time cannot be arbitrarily reduced~\cite{beauchardSCL2014,beauchardMMAS2018}. The same conclusion was obtained in~\cite{beschastnyiJMP2021} for a large class of systems using semiclassical analysis.
More precisly, under mild hypotheses, in \cite{beschastnyiJMP2021} it was proved that for systems which are the quantization of  classical systems with an Hamiltonian containing kinetic and potential energy and controlled via the amplitude of another potential,  the existence of a speed limit  passes through the quantization procedure.

Very recently, it was shown there are  systems relevant for applications for which the controllability time can be reduced to zero between certain states~\cite{duca2021bilinear}. Such systems include planar rotating molecules driven by two electric fields. The results of  \cite{duca2021bilinear} are remarkable as they also  apply to the nonlinear Schr\"odinger equation.

\subsection{Goals and challenges for advancing the controllability of open quantum systems}
\label{sec:Theory-Goals}

An overview over the recent literature reveals that for both infinite-dimensional systems as well as finite-dimensional open quantum systems, controllability remains an open challenge~\cite{kochJPCM2016}. At the level of methods, extending finite-dimensional Lie-group and Lie-semigroup techniques to infinite dimensions~\cite{neeb2006towards} is a challenging desideratum. In particular, finding controllability~\cite{keyl18InfLie}
and accessibility conditions as powerful as the Lie-algebra rank condition in finite dimensions would be significant progress. 

At the level of specific open challenges, the question how reachability differs for non-Markovian compared to Markovian dynamics remains largely unexplored. This is of practical relevance to quantum engineering with non-Markovian experimental setups where the  experiment is only sensitive to correlation effects in finite specified time windows. A better understanding of reachability under Markovian and non-Markovian dynamics would allow for estimating the error up to which the  non-Markovian dynamics can still be approximated by a simpler Markovian model.

\subsubsection*{From reachable sets to resource theory}
The characterization of reachable sets \cite{dirr2019,MTNS20}
has so far been given in terms of convex sets containing them. It is a worthwhile next step to distinguish clearly non-reachable states from the convex hull embracing the reachable ones which generically form non-convex sets themselves. This may be possible via the Hahn-Banach separation theorem~\cite{ReedSimonI,Rudin91}.
In analogy to entanglement witnesses, one seeks for linear functionals that give negative values for clearly non-reachable states and positive ones for the convex hull embracing the reachable ones. A further road to specify reachable sets for Markovian dynamics is by devising efficient algorithms for reachability sets as Lie-semigroup orbits when the generating Lie wedge can be given~\cite{ODS11}. Finally, in analogy to the recent
description of a ``distance to uncontrollability''~\cite{burgarth2020}
in closed systems based on earlier symmetry results~\cite{ZS11,ZZS+15}, a feasible measure may also be devised for open systems.

Another obvious goal at the interface between quantum control and resource theory is to generalize the recent reachability results from toy models with  diagonal states (i.e.~states that are diagonal in the eigenbasis of the drift Hamiltonian) to general states. The $d$-majorisation techniques then have to be pushed to the more challenging operator lift of \mbox{$D$-majorisation}~\cite{vEnde-DMaj20,MTNS20b}. At the same time, further interconnections to thermal operations~\cite{Horodecki13,Brandao15,Lostaglio18} and resource theory~\cite{Lostaglio19r} will emerge
by quantifying benefits and limits of heat-bath coupling~\cite{alhambra2019heat} in terms of reachable sets. 
In particular, further insight may be inspired by elucidating the connection between thermal operations and Markovianity~\cite{Lostaglio17,Lostaglio21,dann2021open}.

\subsubsection*{Simultaneous controllability}
While simultaneous controllability can be achieved for a large class of systems depending on one unknown parameter and driven by at least two controls \cite{Agrachev16,beauchardCMP2010,likhaneja2009,augierJCO2018}, it is not clear when this can be obtained via one control only, which however is the most common situation in experimental settings.
For a qubit driven by one control with a dispersion in the Larmor frequency it is possible \cite{robin2020compatibility}. Whether this feature is limited to two-level systems or whether it can be extended to a larger class of systems is an important open problem.  More generally, the role of the drift Hamiltonian in ensemble control is not well understood. While in finite dimensions
(almost) periodicity of unitary one parameter groups allows one to use the forward and backward evolution of the drift Hamiltonian to control a quantum system, it remains an open problem in infinite dimensions.

Another challenge arises from the number of parameters involved.
While systems
depending on two or more unknown parameters are in principle simultaneously controllable~\cite{Chen19} by finitely many independently addressable control Hamiltonians, the minimal number of controls
 -- important for reasonable technical implementations -- is unknown. Furthermore, effective numerical algorithms and in particular reliable error estimates for the implemented approximations are also missing.

\subsubsection*{Quantum speed limit}
For practical applications beyond two-level and three-level systems, quantum speed limits have so far been determined numerically in heuristic way~\cite{canevaPRL2009,goerzJPB2011,goerzNPJQI2017}. It will be interesting to see whether recently introduced alternatives for determining the {\sc qsl}, for example in terms of semidefinite programs~\cite{zhangPRL2021} or the changing rate of phase~\cite{sunPRL2021}, provide a more systematic and numerically less costly approach. Another promising development concerns use of the {\sc qsl} to assess properties of the system, e.g., reachability of states in an open quantum system~\cite{kobayashi2021}, or properties of the controlled dynamics, e.g., robustness~\cite{kobayashiPRA2020}. This suggests to exploit the various formulations of the  {\sc qsl} in optimization functionals in order to guide a numerical search to solutions with desired properties such as robustness.

While progress on quantum speed limits for open quantum systems in general~\cite{deffnerJPA2017,poggiAFA2020} and their relation to non-Markovianity in particular~\cite{teittinenNJP2019,teittinenEntropy2021} has been remarkable, their identification in many-body systems continues to be an open challenge. Here, the  {\sc qsl} is intimately connected to controllability. The latter may be lost in the thermodynamic limit~\cite{bukovPRX2018}. In fact, to date, for coherently evolving many-body systems, thermal initial states are the only states for which the controllability time has been shown \textit{not} to diverge~\cite{ilinPRA2021}. If confirmed, this result implies that driven dissipative evolution~\cite{verstraeteNatPhys2009} will be the only generally viable route towards many-body quantum control. While it is in line with physical intuition that controlling a many-body system requires simultaneous cooling, both a  thermodynamic and a rigorous mathematical underpinning of this conjecture would be desirable. Recent insight into symmetry classes of open many-body quantum systems~\cite{altlandPRX2021} may provide  guiding principles towards a better understanding of many-body controllability. 

Concerning  systems evolving in an infinite-dimensional Hilbert space (besides the results \cite{boussaidIEEECDC2012,duca2021bilinear,beauchardSCL2014,beauchardMMAS2018,beschastnyiJMP2021}  that treat specific situations), it is not yet entirely clear which properties of a quantum system  imply the non-existence of a quantum speed limit.
Whether and how the existence of a  speed limit
passes through quantization and how this is related to the emerging classicality of certain observables in the spirit of 
\cite{poggiPRXQ2021} are intimately connected problems, which deserve to be investigated.

\subsubsection*{Further characterization of Markovian and non-Markovian quantum maps}
As outlined in Sec.~\ref{sec:Markovianity}, Markovian dynamics is particularly amenable to the framework of bilinear control systems. For practical applications, one first has to assess whether an experiment fits to a Markovian model. Remarkably, given experimental data, the corresponding decision problem (encompassing both the time-independent and time-dependent notion via {\em infinitesimal} {\sc cp}-divisibility~\cite{Wolf08a} on all time scales)  is {\sc np}~hard \cite{Wolf12a,Wolf12}. For a quantum engineer, however, deciding approximate Markovianity on a quantifiable level in the sense 
of ``sufficiently" separate time scales of system and environment dynamics would do in many applications. 

Recently, it was elucidated that time-{\em in}\/dependent non-Markovian refocussing effects may root in correlations on long timescales that appear hidden to observations~\cite{Burgarth-hidden-21} or interventions~\cite{BurgarthModi-2021} on shorter timescales. This, in turn,  paves the way to particularly easy fitting noise of models to tomography data~\cite{Cubitt-Fitting2021}. Exploring whether similar properties hold in the more general case of time-{\em de}pendent memory effects would be helpful. In that case, simpler (time-dependent) Markovian models could serve---within certain time frames---as viable descriptions of dynamics that outside these time windows are time-dependent non-Markovian. Given the complication in non-Markovian control due to memory kernels (see above),
the gain of simplification by a Markovian substitute covering the pertinent time-window of the experiment would be most welcome.
Compatibility with thermodynamics can add an additional aspect to classify the equations of motion in either the
Markovian or non-Markovian case \cite{dann2021non}, cf.~Subsec.~\ref{subsec:quthermo}.

Another route towards approximate descriptions that will ease the control analysis is motivated by a recent example~\cite{trushechkinPRA2021} describing the scenario of a two-qubit system coupled to
a fermionic bath. This can be treated beyond the secular approximation, where on longer time scales the Redfield equation captures the dynamics more precisely than an adapted {\sc gksl}-equation (which on short timescales ensures positivity that the Redfield equation notoriously cannot). 
So finding quantitative guidelines for a sweet-spot in time where to switch from the adapted {\sc gksl}~model to the Redfield model such as to merge the best of the two worlds would allow for obtaining more precise controls in the long-term part of the open system dynamics.

\section{Optimal control methods}\label{sec:methods}
Typically one distinguishes open-loop control where no experimental feedback is used for deriving the control and feedback control. Here, we focus on open-loop control methods which make up the majority of quantum control protocols to date but point out that a combination of optimal control and feedback control has recently been proposed~\cite{Marquardt-GRAPE2022}. 
Different approaches have been proposed to optimize control pulses in the open-loop configuration. Optimal control is born in its modern version with the Pontryagin maximum principle (PMP) in the late 1950s and applied to quantum dynamics since the eighties. We refer the interest reader to the recent review~\cite{boscainPRXQ2021} for an in-depth mathematical introduction. Quantum optimal control has then undergone rapid development with a wide variety of methods extending from analytical tools to different numerical algorithms. The analytic approach allows a complete geometric understanding of the control problem from which one can deduce the structure of the optimal solution and, in some cases, a proof of its global optimality. Physical limits of a given process such as the minimum time to achieve a state to state transfer can also be derived. We stress that such results can be determined analytically or at least with a very high numerical precision. 

The numerical approach is generally based on algorithms which compute iteratively control processes that are closer and closer to the optimal solution. This method has key advantages which are complementary to the ones of analytic computation. They are first applicable in complex quantum systems, that is not the case of analytic tools which can be used only for quite simple systems. The flexibility of numerical algorithms makes it possible to adapt them to specific control problems or to experimental limitations and uncertainties. This aspect is crucial to fill the gap between theory and experiments.

Note that different families of numerical algorithms have been developed according to the characteristics of the optimization problem. Among others, we can distinguish the size of the system, the precision of the optimization process, the type of constraints on the state or the control or the figure of merit to maximize. Such  algorithms can be roughly divided into two groups, namely the gradient-based numerical methods and the gradient-free ones~\cite{glaserEPJD2015}. As their names suggest, the update of the control sequence is either based on the calculation of the gradient of the figure of merit or on a direct search method, i.e. without gradient. More precisely, when the pulse sequence is parameterized by piecewise constant controls in time, there are two well-established gradient-based optimal algorithms, the Krotov-type~\cite{ReichJCP2012} and the GRAPE-type (Gradient Ascent Pulse Engineering) methods~\cite{KHANEJAJMR2005}. The main difference between the two approaches concerns the update of the control which is sequential in Krotov schemes, while is concurrent in GRAPE procedures. A systematic comparison and a discussion of the relative advantages of the two optimization processes can be found in~\cite{machnesPRA2011}. 

An alternative route is that to expand the pulse sequence in a functional basis,  e.g. a Fourier basis, and considering only a limited number of basis functions to represent the control. This approach is justified by the exponential convergence the achievable precision with the number of basis functions~\cite{Lloyd2014a}, that allows to drastically reduce the optimization problem complexity. Moreover, a randomization of the basis function improves convergence properties of the optimization. Within this approach, which goes under the name of CRAB (Chopped random-basis quantum optimization)~\cite{canevaPRA2011,doriaPRL2011,Rach2015}, gradient-free approaches are an interesting and efficient alternative. In this framework, the optimization problem can be transformed from a functional one to a multi-variable optimization that can be solved with a direct-search method. The development and current status of CRAB  has been recently reviewed in~\cite{mueller2021}.

Gradient-based and gradient-free algorithms are two complementary methods with their relative advantages and limitations. A very interesting aspect of gradient-free approaches is their simple way of being implemented both numerically and to take into account experimental constraints. They can also be used for controlling high-dimensional quantum systems. On the other hand, their precision and the type of controls that can be generated are limited by construction. Such problems can be overcome by gradient-based methods, the price to pay being a higher numerical cost and a more technical and mathematical implementation. In this direction, the efficiency of an exact-gradient based optimal control methodology \cite{Fouquieres2011} has recently also been shown in the control of many-body systems~\cite{jensenPRA2021}. 
In experimental settings with only a limited number of control amplitudes, discrete-valued-pulse optimal control algorithms can be useful \cite{Dridi2015}.

Several software packages for QOCT implementing the methods listed above are by now publicly available. These include software for the Krotov approach for both unitary and open system quantum dynamics~\cite{goerzSciPost2019} and for quantum circuit optimization~\cite{li2021}. Simulation and control of spin systems have been developed in Spinach~\cite{spinach2011}. An  open-source code specifically designed to solve quantum control problems in large open quantum systems whose dynamics are governed by the GKLS master equation is described in~\cite{guenterQuandary}. A software framework for simulating qubit dynamics and robust quantum optimal control is proposed in~\cite{teske2021} with a special emphasis on simulating realistic noise characteristics and experimental constraints. Note that these methods can also be combined with software for computing quantum dynamics~\cite{johanssonCPC2013}. Software tools are described in~\cite{ballQST2021} to help with the application and integration of quantum control in the framework of quantum computing. A description of the different steps going from control, characterization to calibration of the use of quantum devices applied to superconducting qubits are discussed in~\cite{wittler2020}.

\subsection{Recent progress on optimal control methods}

We review in the section the recent progress done in the development of optimal control methods since Ref.~\cite{glaserEPJD2015}.

\subsubsection{Analytic approach}
A series of fundamental and practical issues in quantum technologies have been solved recently using analytical techniques. These studies consider benchmark control problems for ideal quantum systems in which some experimental limitations are neglected. Illustrative recent examples are the optimal synthesis of SU(2) operations on a single qubit~\cite{albertini2015,garon2013}, state control in a spin chain~\cite{stefanatosPRA2019,albertiniIEEE2018}, or the simultaneous control of two or more uncoupled spins~\cite{romanoJPA2016,albertini2016,assematPRA2010,jiPRA2018,dAlessandro2019}. 

Even more difficult control processes such as the design of robust or selective pulses with respect to parameters of the system Hamiltonian have been explored. The basic idea consists in simultaneously controlling an ensemble of identical systems, here qubits, that differ only by the value of an unknown parameter.  Robust, respectively selective, optimal pulses correspond to the case where the target states are the same, or different. In the simplest case with very few qubits, robust pulses can be derived exactly~\cite{vandammePRA2017b,dridiPRL2020}. More complex systems require approximations such as linearization of the dynamics~\cite{martikyanPRA2020b,linaturecomm2017} or a perturbative expansion~\cite{vanDammePRA2021,dongPRXQ2021,buterakosPRXQ2021,ZengNJP2018,zengPRA2018,zengPRA2019}. Control protocols to enhance selectivity or discriminative power have been derived for both state-to-state transfers and unitary transformations~\cite{vanDammePRA2018,ansel2020,basilewitschPRR2020}. Another example are entangling operations for two qubits where the control problem can be mapped to geodesics after separating local and non-local contributions to the evolution~\cite{tang2022}.
An alternative route to robustness are generalizations of adiabatic evolutions such as the Derivative Removal by Adiabatic Gates (DRAG) framework~\cite{Motzoi2009} which has recently been extended to $\Lambda$-systems~\cite{vezvaee2022}.

The analytical approach is not limited to closed quantum system, but can also be applied to open quantum dynamics. The optimal synthesis of a qubit interacting with a Markovian bath can be completely derived~\cite{linPRA2020,lapertPRL2010}. Relaxation-free subspaces for perfect state transfer in $N$-level systems with finite-power are obtained  if and only if each decaying state is connected to two non-decaying states~\cite{yuanPRA2012}. The physical limits for fast qubit reset, where the qubit interacts with a structured environment consisting of a strongly coupled two-level defect and a thermal bath, have been derived in terms of minimal time and maximum purity~\cite{fischerPRA2019,basilewitschNJP2017,basilewitschPRR2021}. 

Analytical techniques and the Pontryagin Maximum Principle may also play a more unexpected role in quantum computing issues where the goal is not to find a time-dependent control, but to optimize quantum algorithms or circuits. It has recently been shown that a standard time-optimal control can be mapped to a Grover’s quantum
search problem~\cite{linPRA2019}. 
A general connection between optimal control theory and variational quantum algorithms has been established~\cite{yangPRX2017}.
Such a link can be used, for example, to precisely adjust the parameters of a quantum circuit~\cite{magannPRXQ2021}. 
Optimal protocols have been derived also for quantum annealing problems~\cite{bradyPRL2021}, attesting to the usefulness of optimal control theory as a general optimization tool in various areas of application in the quantum technologies.

\subsubsection{Numerical approach}
\label{subsec:num_methods}
Major progress has been made with optimization algorithms, ranging from their numerical implementation, adaptations of these algorithms to specific problems that arise in the quantum technologies all the way to the role of measurement for the pulse design. We start be reviewing the progress that has been made in the numerical implementation of optimal control algorithms. It is important for studying increasingly complex systems and for taking experimental limitations and uncertainties into account. 

In terms of better numerical efficiency, time parallelization accelerates the execution of quantum optimal control algorithms~\cite{riahiPRA2016}. Memory requirements in GRAPE can be reduced by exploiting the fact that the inverse of a unitary matrix is its conjugate transpose in combination with automatic differentiation~\cite{narayanan2022}.
A modified version of GRAPE, based on a Krylov approximation of the matrix exponential, allows for dealing with high-dimensional Hilbert spaces~\cite{laroccaPRA2021}.
A global optimization algorithm with quantics tensor trains has been proposed~\cite{soley2021}. 
Improved convergence is obtained when including second order derivative information. For example, a Newton-Raphson method with a regularized Hessian can be applied~\cite{goodwinJCP2016}. This should be used on top of exact, yet efficient calculations of the gradient~\cite{goodwinJCP2015,jensen2020b} since approximations of the gradient also limit convergence of gradient-based methods. For spin systems in particular, the su(2) algebra can be exploited to calculate both first and second derivatives exactly~\cite{foroozandehAutom2021}.
Similarly, faster optimization is possible by generalizing Krotov's method to second order in all derivatives~\cite{shao2021}. 
Further speed-ups are possible by replacing standard time propagation with a product of short-time propagators~\cite{dalgaard2021b}. 

The topology of quantum control problems that governs the convergence of the optimization algorithms, often termed control landscape, has recently been reviewed~\cite{Ge2022}. Analysis of the quantum control landscape can be exploited to derive the Pontryagin maximum principle for robust control~\cite{koswaraPRA2021,koswaraNJP2021}. Robustness comes, however, at the expense of solving partial differential equations for the time evolution of the system~\cite{koswaraPRA2021,koswaraNJP2021}.
Analytic descent for parameter optimizations can approximate the corresponding control landscape locally~\cite{koczor2022}.
Optimal control solutions can be reshaped and guided to produce user-customized solutions by using the geometry of control landscapes~\cite{laroccaPRA2020a,laroccaPRA2020b}. 
Improved performance of gradient-based quantum control algorithms has also been found using {\it push-pull} optimization, where a hybrid cost function is used to maximize the overlap with the desired target state while minimizing the overlap with orthogonal states \cite{Batra2020}. This is closely related to optimizing a target functional while penalizing the population of undesired states via time-dependent targets~\cite{palaoPRA2008,ReichJCP2012}.  

Another difficulty in optimal algorithms is how to initialize the  guess field. A recent study has put forward from a citizen science game the hypothesis that human common sense could help the algorithm in this first step~\cite{jensenPRR2021}. 

In open quantum systems, quantum trajectories instead of the density matrix formalism and automatic differentiation have been used~\cite{leungPRA2017,goerzQST2018,leungPRL2018} to speed up  optimization and reduce the computation complexity. 
Steady states of dissipative dynamics including non-equilibrium states can be targeted by optimal control via implicit differentiation~\cite{vargas-hernandez2020,vargasHernandez2021}.
Optimal control has been combined with time-convolutionless master equations~\cite{basilewitschNJP2019,wuPRA2022} which allowed for studying, respectively, qubit reset and instantaneous tracking under non-Markovian dynamics. Such a combination allows for investigating the impact of control on dissipators~\cite{basilewitschNJP2019,kallush2022}, as a means to implement QOCT for quantum reservoir engineering~\cite{hornNJP2018}. The simplest way to treat memory effects consists in employing structured environments that are partitioned into strongly and weakly coupled modes~\cite{ReichSciRep15,basilewitschNJP2017,fischerPRA2019,basilewitschPRR2021,ansel2022}. Novel approaches to Non-Markovian dynamics that allow for a more detailed description of condensed phase environments have also been combined with QOCT~\cite{mangaudNJP2018,fuxPRL2021}. 

Different approaches have been introduced to better account for experimental limitations. For example, gradient optimization of analytic controls (GOAT) is a new algorithm that allows to target high fidelities while designing pulses that conform to hardware-specific constraints~\cite{machnesPRL2018}.
B-splines can be used for pulse parametrization in gradient-based optimization~\cite{guenther2021,peterson2021,ozguler2022}. A binary relaxed gradient in which the pulse is either on or off has been introduced~\cite{vogt2021} for generating unitary gate transformations. Multiple constraints in gradient optimization can be accounted for via auto-differentiation in Tensorflow~\cite{songPRA2022}. High-efficiency control sequences compatible with experimental constraints can also be designed based
on the Magnus expansion where the corrections necessary to reach high fidelity are found order by order~\cite{roqueNPJQI2021}.
Riemannian optimization techniques for solving constrained optimization problems are proposed~\cite{luchnikovSciPost2021} for quantum technology applications. GRAPE can be modified to include binary control pulse optimization~\cite{fei2022}. Time-correlated multiplicative control noise can be mitigated based on a circuit-level representation
of the control dynamics~\cite{trout2022}.
Also, gradient-free optimal control can be formulated such as to yield phase modulated-only driving fields which are more robust than pulses which are both amplitude and phase modulated~\cite{tianPRA2020}. 

The applicability of existing numerical algorithms to the quantum technologies has been improved by tailoring to specific tasks such as system identification~\cite{Brockett2001,NdongJPA2014,wangIEEE2018,buchwaldPRA2021,anselPRA2017, Asslaender2017}.
Based on a specifically tailored target functional, robust control sequences have been optimized to measure rates of stochastic processes \cite{Nguyen2017}.
A number of target functionals specifically adapted to quantum technologies have been introduced, for example to maximize the quantum Fisher information for quantum metrology~\cite{linPRA2021,linPRA2022}, to optimize for mixed target states often encountered in squeezing~\cite{basilewitschAQT2019}, to design swap operations for encoded qubits~\cite{basilewitsch2021}, or to optimize transport~\cite{cole2021}. Several target functionals have been tested in quantum estimation~\cite{basilewitschPRR2021,laverick2021}.
An algorithm for periodic quantum dynamics has been proposed to maximize the signal to noise ratio~\cite{jbiliPRA2019}. In order to find optimal control solutions in  presence of large measurement shot noise, use of Bayesian inference has been suggested~\cite{sauvagePRXQ2020} and the efficiency of a modified gradient-based approach which allows for feedback to stochastic quantum measurements has been demonstrated on a Jaynes-Cummings model \cite{Marquardt-GRAPE2022}. 
Robustness against specified frequency bands of the noise power spectral density can be achieved based by including a filter function which can either be derived from reverse engineering~\cite{colmenar2022} or parametrized and optimized~\cite{yang2022}.
The design of robust control protocols has been explored for different sources of imperfections~\cite{schirmer2020,tabatabaei2020,le2021,propson2021,barr2022} and the optimal control of an inhomogeneous spin ensemble coupled to a cavity has been studied \cite{Ansel2018}.

Numerical optimal control algorithms are based on an open-loop configuration in which no feedback from the experiment is used during the control process. This type of control is obviously limited by the precision of the modeling. The next generation of algorithms will have to take measurement data into account in the design of the control protocol~\cite{ZHANGPhysRep2017}. A first step in this direction is to explore how optimal control and the design of the corresponding pulses can be assisted by experimental data, such as von Neumann measurements~\cite{sugnyPRA2008,shuang2008PRA,pechenPRA2015}. These data may modify the characteristics of the optimal solution and its construction~\cite{muleroJMP2020}. 
In this setting, the ability to control and reconstruct the full state of the system from different measures has been explored~\cite{arenzPRA2020}. But also single qubit measurements alone are useful to assist quantum control~\cite{policharla2020}.
A theoretical framework combining a resource-efficient characterization and control of non-Markovian open quantum systems has been developed~\cite{chalermpusitarakPRXQ2021}. Standard approaches of direct feedback control in which a function directly proportional to the output signal is applied have been extended and applied to quantum control problems~\cite{chenNJP2020}. A data-driven regression procedure that leverages time-series measurements to establish quantum system identification for quantum optimal control has been proposed~\cite{goldschmidtNJP2021}. Further proposals to combine machine learning approaches and QOCT will be reviewed below in Sec.~\ref{subsec:QOCTvsML}.

\subsection{Goals and challenges for advancing  optimal control methods}
Controlling quantum systems with high efficiency in minimum time is highly important for quantum computing and more generally quantum technologies. Control laws are generally computed analytically or numerically in an open-loop fashion on the basis of a theoretical model of the dynamical system. In this setting, a mid-term objective is to continue the numerical development of both gradient-based and gradient-free algorithms with the aim of treating increasingly complex quantum systems and accounting for all relevant experimental details while keeping calculation times feasible.

In spite of recent progress, a main obstruction to the experimental realization of optimized control pulses remains their high sensitivity to experimental imperfections and model uncertainties. This problem has motivated the development of methods addressing control robustness since the early days of the field but important  limitations remain. For instance, it is currently not possible to mitigate random fluctuations, and robust controls are very system-dependent. A general protocol for the design of robust pulses against stochastic variations of system parameters would be a crucial step forward. 

Another key objective is to leverage optimal control techniques for a better characterization of quantum systems and thus enhance the accuracy of the used models. One idea is to use optimally shaped pulses for generating a transformation that maximizes the difference in system response to different parameter values, in order to facilitate their measurement. To date, this approach has been tested on small model systems, and a mid-term goal would be to identify the most scalable version among these approaches.
In principle, such methods allow for a full characterization of both the system and its environment. They are thus of interest in different quantum technology applications, and possibly offer new directions in quantum metrology and quantum sensing. 

On the other hand, the control of macroscopic systems in robotics or mechanics is very often carried out in a closed loop scheme with a only basic knowledge of the system dynamics. Real-time measurements allow the operator to correct and systematically adjust the system trajectory in order to reach the desired target or to carry out the expected task. While this approach to control is the most efficient way to manipulate a system in a way that is robust against any form of disturbance, it is difficult to transfer it to the quantum world due to the cost associated with measurements and the short timescales of quantum dynamics. However, various recent technological advances give hope that this objective is not out of reach. The main objective of the next generation of quantum optimal algorithms will be to take measurement data into account in the optimization procedure, the ultimate goal being to achieve a quantum computation controlled in real time.

\section{Similarities and differences between QOCT and related approaches}\label{sec:comparison}

\subsection{Closed-loop vs open-loop control} 
Quantum optimal control as defined in this review is assuming that the time evolution of the quantum system is not actively observed by the controller during the time span of the control. It could still be an open system evolution but the information imprinted in the environment is not being used. This differentiates quantum optimal control from quantum feedback \cite{wiseman2010}, where information is extracted and used to construct feedback controls in real time. Both have advantages and disadvantages in their own right. 

Still, within the domain of quantum optimal control in this sense, one can discriminate open-loop and closed-loop approaches, which in the more modern language of inference and learning can be called offline and online methods respectively. In the open-loop / offline approach, a model of a physical system (e.g. a Hamiltonian or the ingredients of a suitable master equation) are used to perform the control calculation and then applied to an experiment that is described by that model. On the other hand, closed-loop / online approaches directly use an experimental setup to perform optimal control instead of a mathematical model, hence in its purest form performing a model-free optimization. 

While the former is described in other sections of the review, it is worth highlighting the ingredients of the latter a bit more. Here, one needs to find an experimentally accessible version of the stopping criterion for the optimization: 
The cost function needs to be {\em measured} instead of being computed, and a rule for pulse updates needs to be formulated. Depending on the isolation of the system, measuring the gradients might be rather imprecise and, depending on the level of noise, gradient-free methods may be a better suited alternative. A simple and robust approach relies on randomized benchmarking for quantum gates and Nelder-Mead optimization \cite{Egger2014closed,kellyPRL2014} or more advanced optimizers \cite{Ferrie2015closed}. 
Model predictive control has been suggested as a closed-loop optimization framework that inherits a natural degree of disturbance rejection by incorporating measurement feedback while utilizing finite-horizon model-based optimizations to control multi-input, multi-output dynamical systems~\cite{Goldschmidt2022}.
Special attention needs to be given to the single-qubit case as measuring requires single qubit gates at least, but this can be bootstrapped \cite{Dobrovitski2010closed}. An impressive experimental test of this approach has been achieved in quantum dots \cite{cerfontaineNatComm2020}. 

Advanced methods combine both approaches on their merit - using the efficient, fast and -- within the model -- precise convergence of open-loop techniques in combinations with closed-loop controls that contain the complete experimental reality and provide data for updating the underlying model~\cite{dalgaardNPJQI2020}. 

Another related approach appears in the area of variational quantum algorithms~\cite{cerezo2021}, believed to be advantageous for noisy NISQ quantum computers~\cite{Preskill2018}, that is rather similar to closed-loop optimal control. In these approaches, a quantum algorithm that contains parameterized gates (i.e., gates that contain, e.g., a rotation angle as a free parameter) is considered with the goal of moving a fiducial initial state into a desired final state. In the language of quantum optimal control, it is a state-transfer problem. In order to find the parameters from an initial {\em ansatz}, the desired figure of merit or cost function is measured in the end of the algorithm and according to its outcome, a classical optimization algorithm updates these parameters. There are two classes of such algorithms: (i) The state preparation consists of a set of sufficiently general operations and does not involve a quantum representation of the cost function - this is, e.g., the case in the variational quantum eigensolver (VQE) \cite{Peruzzo2014} for theoretical chemistry \cite{Kuehn2019}, high energy physics \cite{kokailNature2019} or computationally hard problems in graph theory~\cite{ebadi2022}. It requires a careful determination of the {\em reachable set} of states and a sufficiently large number of controls in order to be able to reach a sufficient approximation of the desired final state. A popular choice is, e.g., the set of unitary coupled cluster states \cite{Romero_2018}. (ii) A so-called "cost Hamiltonian", encoding the desired solution to the optimization problem, and a fairly simple driver Hamiltonian, also referred to as mixer Hamiltonian which does 
not commute with the cost Hamiltonian are alternated for adjustable durations, as is the case for the QAOA algorithm \cite{farhi2014quantum} (Quantum Alternating Operator Ansatz or Quantum Approximate Optimization Algorithm). While this has been proven to be universal, clear proofs of quantum acceleration are difficult \cite{hastings2019classical,dalzellQuantum2020}. In both of these cases, the controls are written as parameterized gates in a quantum circuit, which can be interpreted as a very simple parameterization of a long control pulse. Recently, this has been taken to the domain of more continuous pulse parameterizations \cite{bradyPRL2021,mbeng2019b,choquettePRR2021,dekeijzer2022}. In both cases, insights of optimal control theory around reachability, speed limits, and required number of parameters apply. Further examples for cross-fertilization between QOCT and variational quantum circuits are reviewed below in Sec.~\ref{subsec:qualg}.

In more practical terms, successful VQEs require fast and reliable classical optimization algorithms. A recent comparison of four different gradient-free optimization methods revealed superior performance of stochastic methods, in particular when used with default parameters~\cite{bonet2021}.

\subsection{Quantum optimal control vs machine learning approaches}
\label{subsec:QOCTvsML}

Machine learning is a field of computer science which has been attracting much attention in many areas in physics~\cite{carleoRMP2019,mehtaPhysRep2019,judsonPRL1992}. The algorithms are built to emulate human intelligence by learning the best way to proceed from a large data set~\cite{mnihNature2015}. The power of this tool gives hope that long outstanding problems can be solved. For example, in quantum physics~\cite{HushScience2017} it has been applied with success in many-body physics~\cite{CarleoScience2017,carrasquillaNatPhys2017} and quantum computing~\cite{carleoRMP2019,dunjkoPRL2016}. We review in this section recent studies investigating quantum optimal control problems.

A specific branch of machine learning, namely reinforcement learning (RL), is generally used to solve such problems. This approach is intimately related to optimal control theory even if the way to describe the optimization process might look different at first glance. In RL, an agent takes actions in order to maximize a current or a final reward. The learning process is based on observing the effect of the action on the dynamical system and on the reward. From this information, the agent decides to modify (or not) the action. Replacing in the previous description, agent by control, action by control law and reward by fidelity, the parallel between RL and QOCT becomes immediate. The connection holds also in terms of the underlying mathematical structure since RL can be viewed as a dynamic programming approach which in turn is based on the Hamilton-Jacobi Bellman formulation of optimal control~\cite{brysonBook}. Finally, the main difference between the two formalisms is the way to determine the new control from the previous ones in an iterative optimization process. In particular, RL is expected to add value to optimal control techniques in the case of a complex control landscape with many local maxima. Indeed, the procedure for designing control law can escape local traps through random changes in control. A short introduction to the different learning control methods in quantum physics is given in Ref.~\cite{dong2020} and Ref.~\cite{giannelli2021} provides a tutorial-style introduction into both optimal control and reinforcement learning.

These ideas have recently been explored in benchmark quantum control problems in order to show the efficiency of this RL approach. In a system of coupled spins with bang-bang controls, RL has been shown to lead to a quasi-optimal fidelity~\cite{bukovPRX2018}. Drawing a parallel with statistical mechanics, this study also interprets the change of structure of the control landscape as a phase transition, highlighting the specific role played by time-optimal control process~\cite{dayPRL2019}. Quantum speed limits can be found in a spin chain~\cite{zhangPRA2018}. Basic questions such as the implementation of quantum gates~\cite{AnEPL2019}, the transport of quantum states~\cite{porottiCommPhys2019} and robust control against leakage or control errors~\cite{niuNPJQI2019,wuPRA2019} have been answered. The efficiency of learning algorithms with experimental feedback has been explored~\cite{yangPRA2020} and the quantum analogue to the classical cartpole balancing  problem has been analyzed~\cite{wangPRL2020}, showing that RL matches or outperforms other methods in this example. A standard limitation of open-loop control protocol is model bias. This limitation can be overcome by RL in which the agent learns the system parameters through a series of interaction, corresponding here to measurements, with the quantum system~\cite{sivak2021}.

Another difficulty in numerical optimal control is the choice of a good guess field to initiate the optimization process. This obstacle can be overcome by a learning approach~\cite{dalgaardNPJQI2020}. A very good modeling of the cost functional landscape can be achieved from deep learning. The efficiency of this approach has been shown for the control of spin chains~\cite{dalgaard2021}. RL has been combined with analytical control pulses for spin manipulation in order to account for robustness constraints~\cite{ding2020}. Another promising example is given by a hybrid algorithm using a quantum computer as an active part of the optimization process, devising the control of a molecule by a laser field~\cite{castaldoPRA2021}: The time evolution of the wave packet is determined from a quantum computer, while the iterative procedure is realized by a machine learning algorithm. A model-based RL is investigated in Ref.~\cite{schaferMLST2020} for different control problems. The authors show that gradient-based approaches can be combined with learning processes to speed up the optimization procedure~\cite{schafer2021}. This approach named differentiable programming is expected to be much more efficient than model-free RL. It has been applied for state preparation and stabilization of a qubit subjected to homodyne detection, in which the qubit dynamics are governed by stochastic Schr\"odinger equation, which is very difficult to deal with using standard optimization methods. A recent systematic comparison has shown for qubit manipulation that RL outperforms standard optimization procedures when the problem is discretized and the space of the action is sufficiently small~\cite{zhangNPJQI2019}. RL can be also efficiently combined with gradient-based optimization procedures for the robust control of spin 1/2 networks against different noise sources~\cite{khalid2021}.

Finally, there has been a lot of cross-fertilization between QOCT and machine learning recently~\cite{dunjkoQV2020}. In particular, 
QOCT can directly be exploited in the design of quantum algorithms and the synthesis of quantum circuits. The corresponding work is reviewed below in Secs.~\ref{subsec:qualg} and~\ref{subsec:qucompil}. 
In turn, the quantum variational agent of a variational quantum circuit can learn to solve the quantum control problem~\cite{sequeira2022}. Similarly, machine learning methods such as the recommender system can be used to expedite both GRAPE and a hybrid method combining GRAPE and simulated annealing~\cite{batra2022}.
As already mentioned, another recent development is to combine GRAPE with feedback making use of reinforcement learning~\cite{Marquardt-GRAPE2022}.
These results, as well as those covered in Secs.~\ref{subsec:qualg} and~\ref{subsec:qucompil}, highlight that QOCT as a general optimization tool is not only interesting for the computation of time-dependent control pulses, but also in other optimization problems of interest in quantum technologies. In this framework, geometric control has been combined with machine learning techniques in order to improve the synthesis of quantum circuits~\cite{perrierNJP2020}.

\subsection{Quantum optimal control vs shortcuts to adiabaticity}\label{subsec:shortcuts}

Shortcuts to adiabaticity (STA) is nowadays a well-established set of control protocols which have recently been reviewed in depth~\cite{gueryodelinRMP2019,deffnerPRX2014,torronteguiAAMOP2013}. In short, STA exploit the algebraic structure of quantum mechanics and correspond to fast
routes between initial and final states that are connected through a slow (adiabatic) time evolution when a control parameter is changed in time. They also aim to preserve as much as possible the robustness of adiabatic dynamics. STA solutions are generally different from optimal ones and provide a complementary strategy which has peculiar advantages and limitations as discussed below. Optimal control pulses are built on a global constraint, the minimization of a cost functional, while STA techniques are primarily built to account for local constraints, in particular at time interval boundaries. These two different ways of attacking a control problem show that the two formalisms can mutually benefit from each other.

Recent work has shown the flexibility of STA which can be applied to a wide spectrum of quantum systems. Such studies focus on  the fast and robust transfer in a Su-Schrieffer-Heeger chain of quantum systems~\cite{angelisPRR2020},
fast and accurate adiabatic quantum computing~\cite{hegadePRApp2021,hegdePRA2022},
the control of Bose-Einstein condensates~\cite{zhuPRA2021,chenPRA2018},
applications in quantum thermodynamics~\cite{hartmannPRR2020,abahPRR2020,funoPRL2017},
the generation of quantum gates~\cite{palmeroPRA2017,wangPRApp2019,yanPRL2019,kang2021}, the experimental initialization of spin dressed states~\cite{kolblPRL2019} for optimal control procedures. In view of the development of quantum technologies, this also includes, among others,
the robust preparation of non-classical states~\cite{abahPRL2020,chenPRL2021} as well as many-body states~\cite{carolan2021}, the slowing down of particles by laser fields~\cite{bartolottaPRA2020}, the propagation of matter waves in curved geometry~\cite{impensPRL2020}, the manipulation of two coupled Harmonic oscillators~\cite{tobalinaPRA2020}, the control of multi-level quantum systems~\cite{liPRA2018,vitanovPRA2020,angelisPRR2020} and the displacement of a trapped ion~\cite{anNatComm2016}. STA has been combined with machine learning techniques in~\cite{banSciRep2021} to speed up the quantum perceptron, a fundamental building block for quantum machine learning. Enhanced shortcuts to adiabaticity have been recently proposed to broaden the scope of the approach~\cite{whittyPRR2020} and STA protocols robust with respect to different sources of noise have been derived~\cite{LevyNJP2018,daemsPRL2013,ZengNJP2018,zengPRA2018,zengPRA2019,throckmortonPRB2019}. 
Similarly to QOCT, the trade-off between speed and energy cost of the control process is a key property~\cite{torronteguiPRA2017,campbellPRL2017}.

A key aspect of STA comes from the fact that the control law can be expressed in most cases analytically which makes it possible to highlight the physical mechanisms on which the control process is based. In comparison to QOCT, this
approach requires the system to satisfy specific algebraic properties~\cite{gueryodelinRMP2019} and therefore does not have the full generality of QOCT. In particular, STA cannot be applied directly to any type of dynamical systems or experimental constraints. This obstacle can be overcome by combining STA protocols with optimal control techniques in a hybrid strategy to benefit from the advantages of both approaches. Indeed, the search for the optimal solution and in particular the choice of the cost functional can be guided by the STA solution. This idea is illustrated by recent studies. It has been applied for deriving and implementing experimentally a one-qubit quantum gate~\cite{vandammePRA2017}, to generate specific states in a chain of coupled spins~\cite{stefanatosPRA2019,BalasubramanianPRA2018} and in a three-level quantum system~\cite{MortensenNJP2018}, for the control of entanglement in bosonic Josephson junctions~\cite{StefanatosNJP2018}, but also in many-body physics~\cite{campbellPRL2015}. The connection between STA and QOCT has been discussed in~\cite{zhangEntropy2021} for standard quantum control examples. By using the optimal trajectory as a guide, the authors show that very precise STA protocols can be achieved. Bang-bang control protocols have been derived from the atomic transport in a moving Harmonic trap by using both STA and the Pontryagin Maximum Principle~\cite{dingPRA2020}. QOCT may be combined with STA to mitigate errors, for example those that result from imperfect implementation of an STA trajectory in the fast shuttling of an ion~\cite{Espinos2022}, or improve effectiveness of counterdiabatic local driving of cold atoms in an optical lattice for annealing, state preparation and population transfer~\cite{cepaite2022}.

Conversely, STA can be used to extract information from numerical optimal trajectories~\cite{zengPRA2019}, which further illustrates the link and complementarity between these two formalisms. Similarly, STA can be leveraged to determine one qubit filter functions which can then be used in the cost functional for optimal control to improve robustness with respect to noise~\cite{colmenar2022}.
Under certain circumstances such as the excitation of a two-level quantum system, quantum dynamics can be well approximated by a linear system. In this framework, very strong similarities exist between STA and QOCT protocols, which differ only by the basis of functions on which the two control laws are expanded~\cite{martikyanPRA2020a,martikyanPRA2020b,gueryPRA2014}. In the case of the robust control of a two-state quantum system against different sources of error, a systematic comparison of several methods extending from adiabatic and STA processes to composite and resonant pulses has been performed in~\cite{torosov2021}. They can be combined with dynamical decoupling in order to generate quantum gates in presence of decoherence effects~\cite{whaites2021}. Robust composite pulses mitigating systematic errors have also recently been developed~\cite{kukita2021}. Finally, it should be pointed out that the DRAG technique invented for superconducting qubits \cite{Motzoi2009,Theis_2018} is based on counter-diabatic driving and thus closely related to STA. 

\section{Applications of  quantum optimal control theory}
\label{sec:applications}

When a first version of this roadmap~\cite{glaserEPJD2015} was drafted, QOCT had evolved from a largely theory-driven field to one where theory and experiment started to cross-fertilize. This development has since continued with a steady growth in the number of examples demonstrating significant performance gains thanks to QOCT. We start in Sec.~\ref{subsec:exp} by highlighting successful implementations of QOCT in the laboratory and pointing out issues that needed to be addressed when integrating QOCT into experiment. The following two subsections are then dedicated to a more comprehensive overview over applications of QOCT, organized according to hardware platforms in Sec.~\ref{subsec:hardware} and according to control tasks in Sec.~\ref{subsec:tasks}. Note that some redundancy is intentional to ease locating relevant references.

\subsection{QOCT in experiment}\label{subsec:exp}

Quantum technology has proven to be an ideal testbed for QOCT. First and foremost, quantum technology's precision requirements require an  exquisite understanding -- even in the presence of uncertainties and fluctuations -- of the physical systems that serve as hardware platforms. This is an excellent starting point for quantum optimal control when compared to other fields where QOCT has been explored, such as chemical reaction dynamics~\cite{glaserEPJD2015}. Early scepticism towards QOCT has given way to ready adoption of its toolbox all across the field of quantum technology. Concerns about feasibility, robustness, or intelligibility, due to often somewhat peculiar pulse shapes obtained with QOCT, have been dispelled by proofs to the contrary. We start by highlighting the key experiments at the core of this development, focussing on examples where QOCT results have been taken straight to the experiment or where QOCT has been interfaced with an experiment. For each experimental platform, a more detailed overview will be provided in Sec.~\ref{subsec:hardware}.

Starting with superconducting circuit platforms, QOCT has been used to prepare logical qubits encoded in bosonic modes~\cite{ofekNature2016} and to implement quantum gates~\cite{heeresNatureComm2017,wuPRL2020,zongPRApp2021,werninghausNPJQI2021,matekole2022}. Moreover, STA methods have been used to demonstrate a reduction in  the operation time scale~\cite{yin2021}, and chirped pulses to encode qubits in donor spins in silicon coupled to a superconducting cavity have been utilized to implement a random access quantum memory~\cite{oSullivan2021}. 

Moving to AMO platforms, optimal control has been used to prepare non-classical states in Rydberg atoms~\cite{omranScience2019,larrouyPRX2020} and electric dipole spin waves in an atomic ensemble~\cite{hePRL2020}.
Quantum brachistochrones between distant states of an atom have been demonstrated~\cite{lamPRX2021}, and a source of double twin-atom beams with splitting ramp has been designed using QOCT~\cite{borselliPRL2021}. 
Optimal control has been a key resource for the simultaneous execution of entangling gates in an ion-trap quantum computer~\cite{figgattNature2019}. 
Robust two-qubit gates have been implemented in an ion chain ~\cite{kangPRApp2021} and using microwave near fields~\cite{zarantonelloPRL2019}. 
A gate-optimizing principle trading small amounts of gate fidelity for substantial savings in power, which, in turn, can be traded for increases in gate speed and/or qubit connectivity have also been demonstrated on a trapped ion quantum computer~\cite{bluemelPRL2021}. 
A closed-loop optimization procedure has been used for fast trapped-ion shuttling~\cite{Sterk2022}, and sideband cooling of ions has been optimized~\cite{rasmusson2021}.
The Heisenberg limit in terms of evolution time has been reached in quantum metrology with a photonic platform~\cite{houPRL2021}.
Optimal control of the quantum trajectory of an optically trapped nanoparticle combined with Kalman filtering has allowed for real-time tracking of the particle motion in phase space~\cite{magrini2021}.

Successful application of the QOCT toolkit to color centers in diamond has been continued, for example, to sense temperature~\cite{konzelmannNJP2018} or to study dynamical symmetries that can arise in topological phases of strongly-driven Floquet systems~\cite{wangCappellaro2021}. 
Concatenated pulses have been applied for an easier observation of the Mollow triplet~\cite{wang2020}. Fast and high-fidelity geometric control of a quantum system on hybrid spin registers in diamond has been realized~\cite{dong2021}. 
Robust pulse sequences and optimized pulse pairs have been used to sense temperature and weak AC magnetic fields while decoupling from environmental noise~\cite{vetter2021}, and optimal control of a nitrogen-vacancy spin ensemble in diamond has resulted in an improved detection of temperature and magnetic field~\cite{Poulsen2021}. 
Provably noise-resilient single-qubit gates have been designed with robust optimal control~\cite{zhang2022}.

A further experimental platform where QOCT methods have been employed are spin qubits in a GaAs double quantum dot where 
single-qubit and two-qubit gates with optimized pulses have been implemented~\cite{yangNatElec2019,cerfontaineNatComm2020,cerfontainePRB2020}.

The successful application of QOCT in the lab has been made possible by advances in the integration of pulse shaping techniques with the respective experimental platforms. The development of hardware and interfaces~\cite{forndiazPRApp2017,bertoldiRSI2020,werninghaus2020b,baum2021,xu2021} in particular has been crucial.
Pulse shaping can also compensate for experimental imperfections. This has been demonstrated for signal distortion in  electron spin resonance spectroscopy~\cite{probstJMR2019}.

\subsection{QOCT tailored to specific quantum hardware}
\label{subsec:hardware}

Applications of QOCT can be classified according to the different hardware platforms for quantum technologies or according to different tasks for control. The latter are typically reflected in the choice of optimization method whereas the former are addressed at the level of the underlying dynamical model for system, controls, and decoherence. We cover both classifications, starting with the hardware platforms and allowing for some redundancy between platforms and control tasks, so that readers might quickly identify the literature relevant to their specific concern. 

\subsubsection{Superconducting circuit based architectures}

Superconducting circuits containing Josephson junctions \cite{Makhlin2001} are one of the currently leading platforms to implement quantum computing. They are using the collective electromagnetic variables of superconducting circuits~\cite{krantzARP2019,kjaergaardARCMP2020}, building on the character of superconductivity as a robust macroscopic quantum phenomenon in its own right \cite{Clarke2018}. Hence, they are completely human-made which offers wide options for engineering but also opens the door for parameter uncertainty. Superconducting qubits have strongly improved in coherence of the last decade based on specific design choices as well as material improvements. The theoretical description of superconducting ciruits resembles that of cavity quantum electrodynamics at optical frequencies \cite{blaisRMP2021}.

QOCT has also emerged as immensely important toolbox for the control of superconducting qubit based quantum systems to reach error levels low enough for error correction and NISQ-type quantum computing applications. However, a direct adaption of numerical open-loop optimization concepts suffers mainly from only partial knowledge of the system parameters and consequent requirement of closed-loop protocols with long latency and measurement times. 

In a few situations, analytic pulses can be constructed without resorting to numerical strategies. A well known example is the DRAG-pulse  for weakly-anharmonic transmon-type qubits \cite{Motzoi2009} and recently developed tools based on shortcut-to-adiabaticity methods  for fluxonium-type superconducting qubits  \cite{setiawanPRXQ2021}. Beyond analytical calculations, numerical optimization schemes have been investigated to prepare entangled states in a minimum amount of time \cite{baoPRA2018} and to find fast pulses for controlled-Z gates of frequency-tunable transmon qubits~\cite{garciaripollPRApplied2020}. It has been found that the larger coupling strength between higher-lying energy levels of weakly anharmonic systems, such as the transmon, can be utilized to shorten quantum gate operations~\cite{ashhab2021}. Moreover, with the flexibility in the design parameters of superconducting qubits not only pulse parameters but even system parameters can be optimised to realize a fast universal set of gates with high efficiency~\cite{goerzNPJQI2017}. A rather common source of error for superconducting qubits is crosstalk, and one option to allow for scalability may be to mitigate $ZZ$-crosstalk via co-optimization of pulses and scheduling~\cite{xie2022}. 

Open-loop control methods have been successfully applied to simple systems comprising only a single well-characterized an-harmonic systems to realize gates in higher-dimensional Hilbert spaces  \cite{wuPRL2020} and to prepare logical qubits encoded in bosonic modes, so called cat-states~\cite{ofekNature2016}. Optimal control methods have been used to implement a universal gate set on a logical qubit encoded in a bosonic mode~\cite{heeresNatureComm2017} and are used in combination with STA methods to control a circuit QED system consisting of two coupled bosonic oscillators and a transmon qubit~\cite{yin2021}. For bosonic qubits also, a hybrid approach combining gradient-free and gradient-based optimization has successfully been used to enhance the entangling operation via an effective beamsplitter interaction~\cite{basilewitsch2021}.
Moreover, local (or Lyapunov)  control methods which only require a single forward time propagation of the system wave function to shape an external pulse -- as used for example to steer chemical reactions -- have
been proposed to construct modulated coupler pulses to swap excitations between fixed-frequency qubits ~\cite{malisPRA2019}. 
A fast nonadiabatic controlled phase gate between two transmon qubits with tuneable coupling has been designed using dynamical invariants of motion~\cite{espinos2022b}, and pulse shapes implementing controlled-phase gates based on drive-amplitude and drive-frequency modulation have been derived with a theoretical framework based on Floquet theory~\cite{dipaolo2022}.

However, in many designed quantum systems such as superconducting quantum circuits or defined quantum-dot systems, the information about the underlying Hamiltonian is typically not complete. For instance, the coupling to spurious modes may cause level shifts that are not included in a genuine qubit description. Moreover, parameters may fluctuate on timescales that are comparable or slower than the typical run-time of an algorithm. Pre-determined pulses found in open-loop schemes may therefore not work and one has to resort to closed-loop schemes that optimize a cost function based on experimental data. To go beyond gradient-based algorithms and open-loop control is therefore to rapidly measure the cost function  for each pulse parametrization. Moreover, the optimization algorithm has to make best use of the available hardware by reducing time-consuming pulse reparametrizations and uploads of wave functions from the control PC to the electronics to a minimum. 

To minimize the number of measurements required to completely characterize the quantum operation, typically a fixed-length randomized benchmarking sequence is used for single and two-qubit gates with a few parameters pulses \cite{kellyPRL2014}. It has been demonstrated that this method can be used to achieve fast $4~\rm{ns}$-long single-qubit pulses using piecewise-constant basis functions. Via a closed-loop protocol both coherent gate errors and leakage can be reduced in a  transmon-type qubit \cite{werninghausNPJQI2021}. Numerical studie show that single qubit gate durations can be reduced even further into the $100~\rm{ps}$ regime, provided that the an-harmonicity can be made large enough \cite{zhu2021}. A 'data-driven' version of the GRAPE algorithm that updates the cost function based on experimentally measured state and process tomography data has been used to experimentally realize a controlled-Z gate with 99\%- fidelity~\cite{zongPRApp2021}. 
 
 Ideally, control, calibration and characterization of a system is performed within a general framework that allows for the efficient characterization of the system, the calibration of pulse parameters and the subsequent control of the system~\cite{wittler2020}. For simple pulse calibration, typically only a few pulse parameters, such as amplitude or duration, are optimized. The response of the system to specific measurement sequences is used to iteratively find optimal parameter values for single-qubit \cite{sheldonPRA2016} or  two-qubit gate operations \cite{sheldonPRA2016b,sundaresanPRXQ2021} involving automatic protocols~\cite{xu2021}.  Based on these methods, gates can be calibrated across a complete quantum processor to reach high quantum volumes \cite{sundaresanPRXQ2021, jurcevicQST2021}.  Of particular interest is the mitigation of errors caused by coherent ZZ-type crosstalk, which can be reduced via optimal control pulses also on large scale systems to guarantee high-fidelity parallel gates ~\cite{winickPRL2021}. While the aim is typically to avoid such longitudinal couplings, by robust control of the unitary evolution, large scale  quantum computing on an array of qubits can be envisaged even with fixed longitudinal qubit-qubit interactions~\cite{Nguyen2021}. Similar techniques can be applied  to robustly create GHZ-states of transmons for quantum sensing~\cite{nguyen2021b}.  
 
While transmon-type qubits are the current workhorse of superconducting qubit quantum processors, protected superconducting qubits with exponentially suppressed sensitivity to external noise are heavily investigated. The reduced sensitivity comes at the expense of reduced control possibilities and the need for complex pulses for the implementation of quantum gates. For so-called $0$--$\ensuremath{\pi}$ qubits \cite{brooksPRA2013,gyenisPRXQ2021} control pulses have been optimized that involve higher qubit levels during gate operation to circumvent the intrinsic protection qubit states given by the disjoint support of the low-lying wavefunctions \cite{abdelhafezPRA2020}. Moreover, robust control techniques have been applied in numerical simulations to fluxonium qubits to mitigate parameter-uncertainty errors \cite{propson2021}.

On the other hand, ultrastrong coupling may allow for faster operations at the expense of increased sensitivity to noise in superconducting circuit QED platforms. Optimal control is a tool ideally suited to identify noise-resilient protocols in this setting, for example to realize fast state transfer with noise protection due to an interplay of the dynamical Casimir effect with cavity losses~\cite{giannelli2022}. Further applications of optimal control methods include the concatenation of pulse sequences into a single pulse\cite{Gokhale2019} to realize efficient NISQ-type algorithms or to control hidden qubits that are controlled and read-out via neighbouring qubits \cite{burgarthPRA2009,pechal2020}.

\subsubsection{Color centers in diamond}
Color centers in diamonds are one of the most successful candidates for implementing different quantum technologies, in particular quantum sensors~\cite{Balasubramanian2008,Neumann2013}. Color centers are defects in the regular crystalline structure of diamonds, where one carbon atom is replaced by a different atomic species or a vacancy. This change in the crystalline structure is reflected in the diamond spectral properties that, as a consequence, might gain a colour -- thus the name coloured centers~\cite{Wrachtrup1993,Doherty2013}. More importantly for the quantum technology perspective, these point-like defects have very important and useful properties for quantum information processing. In particular, they satisfy -- to some extent -- the DiVincenzo criteria for quantum computation: it is possible to individuate well defined energy levels to encode qubits states that can be initialized and manipulated via laser excitation and microwaves; the system state can be measured via fluorescence; and different centers can be coupled and entangled, enabling -- in principle -- universal computation and simulation~\cite{Dolde2013,Cai2013,Casanova2016}. Despite these exciting properties, the scalability of the system is still a challenging aspect that, to date, partly limits the applications of this architecture for quantum computation. However, other unique properties make this system ideal to accomplish quantum sensing tasks~\cite{Radtke2019,Barry2020,remboldAVSQS2020}. Indeed, the defect quantum levels are magnetically sensitive exhibiting Zeeman splitting, the quantum properties are stable in a wide range of temperatures, from cryogenic to room temperature and diamonds are biologically inert~\cite{Barry2020}. Thus, in the last years, a very fast development of diamond-based magnetometers for nanoscale sensing also in biological and living cells have been explored and achieved~\cite{konzelmannNJP2018,chakraborty2021,takouPRB2021,whaites2021,marshall2021,hernandez2021,Kucsko2013,Fu2007}. 
Moreover, the sensitivity of the defect properties to strain, electric fields and temperature open the way to a complete new set of sensors of pressure, fields and temperature at the nanoscale~\cite{remboldAVSQS2020}. 

Despite the aforementioned desirable properties of color centers in diamonds, the very same characteristics make them naturally prone to static and dynamical errors, calibration problems, and highly sensitive to drifts of the environmental conditions when implementing, e.g., a spin echo experiment to measure an external magnetic field. Thus, color centers are a natural playground for quantum optimal control, needed to improve the sensing protocols, their stability or final fidelity~\cite{remboldAVSQS2020,Poggiali2018}. Indeed, a number of theoretical and experimental results have been achieved by exploiting the successful interplay between optimal control and color centers in diamonds that span a number of such possible applications. 
Among others, the effectiveness of shaped pulses for temperature sensing~\cite{konzelmannNJP2018}, for initialization~\cite{chakraborty2021}, and for the single-qubit rotations of the electron spin qubit in silicon-vacancy and tin-vacancy defects in diamond~\cite{takouPRB2021} have been demonstrated. On similar ground, experiments have succesfully probed Floquet states for robust control of nuclear spins in NV centers~\cite{whaites2021}, and an optimal two-step approach has been used to improve spin manipulation processes for robust magnetometry with single NV centers~\cite{oshnik2021}. Building on these and other demonstrations of optimally controlled quantum sensing protocols, more complex and challenging protocols have been proposed such as the autonomous calibration of single NV center operations~\cite{Frank2017} and the enhancement of the macroscopic hyperpolarization~\cite{marshall2021}. Finally, interesting links with many-body theory have been unveiled, such as the determination that for a spin sensor of time-varying fields with dephasing noise, the optimal control problem of finding the optimal driving can be mapped to the search of the ground state of a spin chain~\cite{hernandez2021}.

\subsubsection{Trapped atoms, ions, and molecules}
\label{subsubsec:atoms_ions}

Quantum computing with trapped ions is a most promising architecture on par with superconducting qubits and operating according to the gate model of quantum computing. Quantum optimal control has been applied to this platform early on, as summarized in our earlier review~\cite{glaserEPJD2015}. More recent advances include individual addressability of qubits thanks to pulse optimization~\cite{choiPRL2014,figgattNature2019} and robust control, i.e., control pulses that perform well in the presence of parameter fluctuations as well as decoherence. Robustness can be achieved numerically, for example for entangling gates~\cite{bentleyAQT2020,pagano2022}, or using parametric control, i.e. applying sinusoidal modulations to the amplitude, frequency, or phase of the pulses~\cite{greenPRL2015,leungPRL2018,zarantonelloPRL2019,milnePRApp2020,arrazolaPRApp2020}. Moreover, a pulse-shaping technique trading small amounts of fidelity for power savings or trading power savings for gate speed has been used to demonstrate a speed-up of two-qubit gates for a given power budget for trapped ions~\cite{bluemelPRL2021,bluemelnpjQI2021}.
Optimal control theory has been employed to improve the dissipative preparation of entangled states of trapped ions~\cite{hornNJP2018}, and the optimized pumping scheme thus identified has recently been implemented in an experiment~\cite{coledan2021}.

Trapped atoms excited to Rydberg states have emerged as a competitive quantum technological platform, most notably for quantum simulation, with key operations demonstrated for hundreds of qubits~\cite{morgadoAVSQS2021}. Quantum optimal control has been used, first theoretically~\cite{cuiQST2017} and later experimentally~\cite{omranScience2019}, to prepare many-body quantum states of Rydberg atoms in optical lattices. In a similar sequence of theoretical prediction~\cite{patschPRA2018} and experimental demonstration~\cite{larrouyPRX2020}, quantum optimal control has been used to prepare non-classical states in single Rydberg atoms. These can serve as highly sensitive probes of external fields. The accurate preparation of non-classical states of trapped Rydberg atoms relevant in quantum sensing has also been suggested using Bayesian optimization techniques~\cite{mukherjeePRL2020}. Single and entangling gates for Rydberg quantum computing have been optimized for enhanced robustness with respect to parameter fluctuations and decoherence~\cite{goerzPRA2014,guoPRA2020,pelegri2021,jandura2022,pagano2022}.

Quantum optimal control of neutral atoms is not limited to the Rydberg platform, it can also be applied to atoms forming a Bose-Einstein condensate (BEC), see~\cite{glaserEPJD2015} for early work. In this setting, theoretical studies have shown how to control BEC through a variation of the magnetic confinement potential. This study can be performed in a one~\cite{borzi2007,jager2014,sorensen2018,hocker2016} and in a  three-dimensional case~\cite{mennemann2015}. Another degree of control can be achieved by trapping a BEC in an optical lattice where the system can be controlled by a phase modulation of the lattice. A shaken-lattice interferometer can be obtained by selecting specific atom momentum states~\cite{potting2001,weidner2017,weidner2018a}. Transport of BEC with atom chips has been optimized~\cite{corgier2018,amriSciRep2019,chen2011,zhang2016} and robust optimized pulses for cold-atom interferometry have been designed~\cite{saywell2018}. Different experimental evidence for the efficiency of optimal control schemes has been provided. It extends from the loading of an atomic gas in an optical array~\cite{rosi2013,zhou2018} and the manipulation of motional states~\cite{frank2016}, in particular for interferometry applications~\cite{frank2014}, to the crossing of a quantum phase transition~\cite{rosi2013,frank2016} and the transfer to the first vibrational excited states~\cite{bucker2013}. Also, a shaken-lattice interferometer has been experimentally realized through a specific phase modulation of the lattice~\cite{weidner2018}. Recent extensions include remote control of a BEC~\cite{heckPNAS2018,laustsen2021} and state preparation of a BEC in an optical lattice~\cite{dupont2021} as highlighted in Fig.~\ref{fig:QCBEC} in Sec.~\ref{sec:intro}.
In the opposite limit of single atoms, a quantum brachistochrone has been utilized for time-optimal transport between distant sites of an optical lattice~\cite{lamPRX2021}. 

While trapped neutral and ionic atoms continue to be at the forefront of experimental quantum technologies, trapped molecules may eventually emerge as a platform offering more versatility. An example testifying to their versatility is the use of trap-induced resonances to implement two-qubit gates with shaped electric fields for ultracold molecules trapped in optical tweezers~\cite{sroczynska2021}. 

\subsubsection{Other platforms}

Finally, we briefly summarize recent progress in further physical platforms for implementing quantum technologies. Spin states in molecules studied by nuclear magnetic resonance (NMR) have been an early proposed platform for quantum computing, and optimal control applied to this platform has been extensively reviewed in~\cite{glaserEPJD2015}. Recent advances include optimized state preparation in a seven-qubit nuclear magnetic resonance system using hybrid quantum-classical approach to quantum optimal control~\cite{liPRL2017} and the near time-optimal preparation of a Bell state where modeling and experiments were operating in tandem~\cite{chenPRA2020}. Using optimal control, also a novel class of refocusing pulses for "delayed spin echoes" was developed \cite{Asslaender2016, Asslaender2017}, with potential applications in the general field of quantum technology. The concept of optimal-control-based {\it cooperative pulses} \cite{Braun2014}, which are able to compensate each others imperfections, has been introduced for the case of two 90$^\circ$ pulses (separated by an arbitrary delay) as used in Ramsey experiments. The outstanding robustness of cooperative 90$^\circ$ pulses was recently experimentally demonstrated in ultra-broadband multidimensional NMR experiments \cite{Asami2018}. The idea of optimal-control-based cooperative pulses was extended to spin echo experiments consisting of a cooperative 90$^\circ$ and 180$^\circ$ pulse pair with excellent robustness with respect to detuning and scaling of the control amplitudes \cite{Kallies2018}. Furthermore, the physical limits of the time-optimal excitation of maximum-quantum coherence was explored for spin systems consisting of up to five coupled spins \cite{Koecher2016}. Another development motivated by solid-state NMR experiments with potential applications in quantum technologies was the demonstration that it is possible to design control sequences that are robust to periodic modulations of the control amplitude with known modulation frequency but unknown amplitude and phase of the modulation \cite{Tosner2018, Tosner2021}. The optimal-control-based tracking of desired spin trajectories has been used to create highly robust heteronuclear decoupling experiments \cite{Neves2009} and  more recently to tailor the detuning-dependent scaling of the  the spectral splitting caused by spin-spin couplings to reduce the dimensionality of heteronuclear correlation experiments \cite{Zhang2016a}.
In addition to the control of nuclear spin systems in NMR, optimal control pulses have been developed for the robust control of electron spins in the field of electron paramagnetic resonance (EPR) spectroscopy, see \cite{Spindler2017} for a review. In systems consisting of both nuclear and electron spins, the time-optimal polarization transfer from an electron spin to a nuclear spin was explored using optimal control techniques \cite{Yuan2015}.

The ideas developed for atoms and ions that are trapped by external fields, cf. Sec.~\ref{subsubsec:atoms_ions}, are easily carried over to ions that are hosted in a molecule or crystal. For example, triply charged lanthanide ions are a popular quantum platform. Pulses to carry out all quantum gate operations have recently been calculated for the example of gadolinium ions~\cite{castro2021}.  Microwave coherent control of the initialization, operation, and readout of the electronic spin state in erbium dopants has been demonstrated~\cite{cova2021}. Closely related to donor-based solid state platforms are quantum dots, where shaped pulses have been derived for single and two-qubit control~\cite{yangNatElec2019,Kanaar2021,hansen2021,tang2022}. And while Majorana-based topological quantum computation is still elusive, their optimal transport has already been studied theoretically~\cite{coopmans2020}.
 
A large variety of experimental systems exists that realize the coupling of nanomechanical or micromechanical motion to a quantized electromagnetic field mode~\cite{aspelmeyerRMP2014}. Coherent control of cooling such mechanical oscillators and coherent control of energy transfer between mechanical modes  has been demonstrated experimentally~\cite{frimmerPRL2016}. Theoretical proposals for an optimized preparation of non-classical states~\cite{basilewitschAQT2019,bergholmQST2019} and optimized feedback for cooling~\cite{ferialdiNJP2019} have been brought forward.

\subsection{Applications to key tasks in the operation of quantum devices}
\label{subsec:tasks}

We now review applications of QOCT according to control targets, respectively tasks, with the exception of quantum sensing and metrology for which we refer the reader to two recently published excellent reviews~\cite{remboldAVSQS2020,liuAQT2021}. We will start by covering progress in QOCT for state preparation, including transport and storage of quantum information, and measurement in Sec.~\ref{subsec:SPAM}, followed by QOCT implementing desired dynamics in Sec.~\ref{subsec:targetdynamics}. New fields of application of QOCT are quantum algorithms (Sec.~\ref{subsec:qualg}), system identification (Sec.~\ref{subsec:sysid}), and quantum compilation (Sec.~\ref{subsec:qucompil}). We will conclude this overview with reviewing the role of QOCT for quantum thermodynamics in Sec.~\ref{subsec:quthermo}.

\subsubsection{State preparation and measurement}
\label{subsec:SPAM}

Tasks for state preparation are ubiquitous in quantum technologies: In quantum sensing and communication, there are numerous tasks around preparing squeezed, cat and GHZ states vs. photons respectively. In quantum simulation, preparation of a complex state is often the very objective of the simulation task. In quantum computing, there is some focus on unitary gates, i.e., rotations of a full (or computational subspace) basis. Methods for state preparation also pertain to measurements. We review below recent examples in which QOCT is applied for the preparation of a specific state, whereas implementing a desired dynamics that concerns more than a single state will be reviewed below in Sec.~\ref{subsec:targetdynamics}.

QOCT has been used in experiments with Rydberg atoms to increase the fidelity in the preparation highly entangled states~\cite{omranScience2019} and long-lived states~\cite{larrouyPRX2020}. It has also allowed to accurately prepare non-classical superposition states that cannot be prepared with reasonable fidelity using standard techniques~\cite{larrouyPRX2020}.
Shaped laser pulses applied to shift a spin wave in momentum space of atomic ensemble with state-dependent geometric phase patterning, in an error-resilient fashion and on timescales much faster than spontaneous emission~\cite{hePRL2020}.  Starting from a Bose-Einstein condensate, correlated pairs of atoms forming a Bell state involving their external degrees of freedom have been created upon excitation of the condensate with shaped RF pulses~\cite{borselliPRL2021}.

Another example for quantum technologically useful states are squeezed states which allow for quantum enhancement of sensing protocols. In cavity optomechanics, for example, where a mechanical resonator is coupled to a microwave or optical cavity, significant squeezing can be obtained without the need of ground state cooling. In this setting, QOCT allows to speed up the preparation of squeezed thermal states by more than two orders of magnitude compared to a protocol with constant drives, requiring only simple pulse modulations that are fully compatible with current experimental
technology~\cite{basilewitschAQT2019}. 
When further coupling the optomechanical system to a qubit, QOCT can exploit the non-linearity thereby introduced to prepare the mechanical oscillator in non-classical states~\cite{bergholmQST2019}.
In harmonic potentials, squeezed thermal states can be generated by a reverse engineering approach~\cite{dupays2020}. Time-dependent controls for spin squeezing in quantum metrology have been designed with reinforcement learning~\cite{tanPRA2021}.
Large spin squeezing for Ramsey interferometry based on an alternating series of one-axis twisting pulses and rotations has also been designed with QOCT~\cite{Carrasco2022}. 
Coherent-state transfer in the ground-state manifold of an NV center spin using a laser can be accelerated with optimal control~\cite{tianPRA2019}.
Optimal control of the harmonic potential which confines a levitated nano-particle leads to a strong delocalization of its center-of-mass motional state which is expected to enhance force sensing~\cite{weissPRL2021}.  
In a many-body setting, QOCT allows for preparing non-Abelian anyons, important for topological quantum computation, on  timescales many orders of magnitude faster than adiabatic adiabatic dynamics~\cite{raii2022}.

In contrast to state preparation that can be achieved with coherent dynamics, tasks such as qubit reset or measurement are intrinsically non-unitary. Optimizing them requires either the use of open system QOCT~\cite{fischerPRA2019,basilewitschNJP2019,basilewitschPRR2020,guenther2021} or a focus on unitary steps that are part of the overall protocol~\cite{Egger2014b,basilewitschNJP2017,eggerPRApplied2018}. A challenge of qubit reset is that the protocol should work irrespective of the initial state. A naive optimization strategy would seek a control that works for a complete basis of Hilbert space but it turns out that a single, specifically chosen density matrix is sufficient to optimize qubit reset~\cite{guenther2021}. Fundamental bounds on qubit reset in terms of maximum fidelity and minimum time have been determined using the paradigm of resetting via an ancillary quantum system~\cite{basilewitschNJP2017,fischerPRA2019,basilewitschPRR2020}, assuming control over the qubit and no control over the ancilla. 
A practical implementation of ground state reset in a superconducting circuit pumps the
excited-state population to a higher excited state with a first pulse and then dumps it into a low-Q transmission-line resonator, serving as lossy environment, which is also used for qubit read-out~\cite{eggerPRApplied2018}. When the coupling with the thermal environment is tunable, QOCT can be used to determine the optimal tuning protocol~\cite{basilewitschNJP2019}. 

State preparation is also at the core of storing quantum information, and QOCT has recently been used to derive protocols for the optimal storage of a single photon by a single intra-cavity atom, achieving the maximal efficiency by partially compensating parasitic losses~\cite{giannelliNJP2018}. Broadband operation of the quantum memory allows for simultaneously realizing high efficiency and high speed which only requires Gaussian pulses with optimally tuned parameters~\cite{shinbrough2020}.
QOCT combined with a coherent spin-exchange interaction arising from random collisions has been used to derive strategies for high-efficiency storage and retrieval of non-classical light, in order to realize quantum memories with noble-gas spins~\cite{katz2020}. 
A gradient-based optimization strategy has been used to design the temporal shape of the laser field driving a quantum transducer for photons between microwave and optical frequencies to mitigate the effects of inhomogeneous broadening~\cite{mishra2020}.

\subsubsection{Implementing desired dynamics}
\label{subsec:targetdynamics}

We refer to desired dynamics as a control target that concerns more than a single state in the target functional.
A most prominent example are quantum gates.
Optimal control can be used to steer a quantum system toward a target state in a time-minimum way, reaching thus the quantum speed limit. For the example of superconducting qubits interacting via a transmission line cavity, QOCT has 
been used to identify the quantum speed limit not only for a single gate but for a complete universal set, i.e., several local plus one entangling gate, and not only for a single choice of system parameters (qubit frequencies and anharmonicities and cavity frequency) but for the complete design landscape~\cite{goerzNPJQI2017}. This comprehensive numerical study was made possible by combining several advances in the method development of QOCT, including an optimization functional targeting an arbitrary perfect entangler~\cite{wattsPRA2015,goerzPRA2015} and a hybrid two-step optimization approach where the result of a gradient-free optimization becomes the initial guess for a gradient-based optimization~\cite{goerzEPJQT2015}. Optimization towards an arbitrary perfect entangler has also allowed to identify the natural entangling gate for two qubits coupled via a cavity~\cite{goerzNPJQI2017}, while two-step optimization combining gradient-free and gradient-based methods has more recently been employed to enhance the beamsplitter interaction between qubits encoded in bosonic modes~\cite{basilewitsch2021}.
QOCT has been used to derive gates that are robust to secular amplitude drifts~\cite{PhysRevA.104.052625}. Robustness of two-qubit gates can also be achieved by interleaving with 
optimized single-qubit rotations which suppress logical and leakage errors~\cite{setser2020}. Another route towards robust gates are composite pulses. For example, it is possible to derive composite pulses that implement single-qubit gates, such as NOT, designed analytically or numerically \cite{Dridi2020}, that are robust against both detuning and scaling of the control field. The design of broadband or narrowband  excitation pulses was also studied, for example using polychromatic pulse trains~\cite{ivanov2022}.
The generation of NOT and CNOT gates with different optimization methods has been compared~\cite{riazQIP2019}.

Specific gate transformations have also been  optimized for qudits, for example encoded in atomic~\cite{omanakuttan2021} or molecular spins~\cite{castro2021}. Another generalization beyond targeting specific gates is to optimize for a continuous family of gates~\cite{sauvage2021,preti2022}, for example as a function of continuous system or gate parameters. The control landscape for phase shift gates has been found to be free of traps~\cite{volkovJPA2021,volkov2022}. 

Optimization of quantum channels is formally closely related to gate optimization, and the cheapest channel that produces prescribed output states for a given set of input states has been determined~\cite{duvenhage2020}.
Quantum secure data transfer has been optimized~\cite{wang2021}, in which the transmitted data is encoded in the pulse shape of a single optical qubit and high fidelity of the encoding and the receiving processes is implemented with appropriate driving pulses. A unitary transformation of an extended receiver as a tool for quantum state restoring has been studied via optimal transfer of quantum states via spin chains~\cite{feldmanPLA2021}.

\subsubsection{Quantum algorithms}
\label{subsec:qualg}

Quantum algorithms benefit from QOCT in a generic way - gates optimized by quantum control have superior fidelity and thus bring real algorithmic performance closer to the ideal one. 
Application of QOCT can go a lot further, however. On the extreme end, adiabatic quantum computing and quantum annealing forgo the notion of gates completely and focus on the preparation of complex ground states using a slow annealing schedule. QOCT has been shown to be able to find optimal annealing schedules~\cite{mbeng2019a,mbeng2019b,bradyPRL2021}.
In some cases, these are based on the Pontryagin Maximum Principle~\cite{bradyPRL2021,venuti2021}
and may include two independent controls~\cite{fernandes2021}. 

Related to quantum annealing is the Quantum Approximate Optimization Algorithm (QAOA) \cite{farhi2014quantum} which has already been touched upon in Sec.~\ref{subsec:QOCTvsML}.
For gate based quantum computers, QAOA takes the Trotterized version of adiabatic quantum computing and then uses a classical optimizer in order to improve the Trotter parameters. The corresponding crossover between adiabatic quantum computing to QAOA has been considered~\cite{bradyPRL2021,venuti2021}. 

The relation of QAOA and other variational algorithms~\cite{cerezo2021} to QOCT is a lot closer, though~\cite{yangPRX2017,choquettePRR2021,haghshenas2020}. These algorithms employ parametrized quantum circuits, i.e., quantum algorithms whose gates depend on continuous parameters that then are optimized by an external classical algorithm in order to extremalize the variational cost function~\cite{magannPRXQ2021,stokesQuantum2020,yangPRX2017,zhouPRX2020}. In other words, these algorithms use classical closed loop optimal control, choosing the decomposition of a unitary evolution into a set of quantum gates as their waveform parameterization~\cite{magannPRXQ2021}. Note that this argument holds in two versions: In the case of variational algorithms for many-body physics, one aims at a set of controls that suffices to match the desired reachable set of states to then sample from. In QAOA, one is interested in maximizing the overlap with computational basis states of minimum cost function and includes the unitary generated by cost function itself into the control set. It is thus a natural idea to replace the gate parameterization of these algorithms by continuous parameterizations.
Specific structures of quantum algorithms can be identified for quantum optimization which is a key aspect in the development of new algorithms~\cite{mcleanPRXQ2021}. A novel class of variational quantum eigensolvers is obtained by combining optimization and measurement processes, leading to advantages in terms of resources and times~\cite{fergusonPRL2021}. 
Insight from QOCT on overcoming barren plateaus may help overcome convergence problems in quantum machine learning~\cite{holmesPRL2021,broers2021} or in variational quantum algorithms~\cite{larocca2021}.
A QOCT landscape analysis has been applied to the combinatorial optimization problem MaxCut~\cite{leePRA2021}. 
Quantum combinatorial optimization without classical optimization can be based on another class of strategies well-known from QOCT, those inspired by Lyapunov control~\cite{magann2021}. 
Grover's quantum search problem can be mapped to a time-optimal control problem, and then described through the Pontryagin Maximum Principle~\cite{linPRA2019}. 
Furthermore, QOCT has been used to optimize drives in quantum algorithms such as QAOA~\cite{niu2022} and quantum simulation, for example of a chiral effective-field theory~\cite{hollandPRA2020} and of an extended Bose-Hubbard model~\cite{kairys2021}. 

\subsubsection{Quantum compilation and circuit synthesis}
\label{subsec:qucompil}

A first example for the usefulness of QOCT to quantum compilation is given by the Deutsch-Jozsa algorithm which has been compiled on both a superconducting-qubit-based and a spin-chain-based processor using control optimization algorithms together with QuTiP~\cite{li2021}. 
Application of QOCT to quantum compilation and circuit synthesis follows similar lines as that for quantum algorithms outlined above. Indeed, the two questions are closely related since variational optimization is often implemented via Trotterization of a desired unitary, thus  representing an important example for quantum compilation. The corresponding gate sequences can be subjected to optimization, for example to minimize the depth of the circuit on noisy quantum processors~\cite{khatriQuantum2019} or to reduce the approximation error~\cite{mansuroglu2021}. 
Use of variational quantum algorithms, instead of more traditional optimization tools, in order to learn pulse parameters of a quantum circuit, has recently been termed variational quantum pulse learning~\cite{liang2022}.
In order to scale to large circuits, a block-by-block optimization framework has been suggested~\cite{wu2020}.
QOCT and trajectory learning can be combined to map the space of potential parameter values of a quantum circuit to the control space and thus obtain continuous classes of gates~\cite{preti2022}. The same idea can be applied to Hamiltonian simulation~\cite{kairys2021}. When the gates in a circuit are parametrized by continuous parameters, optimization may be hampered by the non-Euclidean nature of the parameter space and proper evaluation of the gradient becomes important, for example via natural gradients~\cite{yao2021}. This problem is reminiscent of the best way to approximate the gradient in QOCT discussed in Sec.~\ref{subsec:num_methods} or the observation that 
estimating the expectation values in hybrid classical-quantum optimization determines the convergences properties of the latter~\cite{SwekeQuantum2020}, highlighting the importance of knowledge transfer between different subfields.

\subsubsection{System identification and calibration}
\label{subsec:sysid}

The identification of parameters that characterize the dynamics of a quantum system is a fundamental prerequisite
for controlling its evolution and realizing concrete tasks in quantum technologies with sufficient precision. It is of particular importance for open-loop configurations in which the control protocols are designed only based on the system model, without any experimental feedback. System identification aims to estimate the value of one or several parameters of the system Hamiltonian. One option consists in building a database from the time evolution of an ensemble of dynamical systems driven by a specific field, which is designed by optimal control theory to maximize the efficiency of the recognition process~\cite{anselPRA2017}. Alternatively, one can use the shape of the driving field to maximize the distinguishability of two states that evolve under slightly different Hamiltonians~\cite{basilewitschPRR2020,qin2022}. System identification is possible even in the presence of large control perturbations~\cite{fu2017}.
When using the quantum Fisher information as figure of merit for estimating the value of an unknown parameter of the Hamiltonian~\cite{linPRA2021}, upper bounds for time-dependent Hamiltonians have been established~\cite{pangNatCom17}. A variational approach combined with shortcuts to adiabaticity has been used to  determine the initial states and the optimal controls that maximize the quantum Fisher information~\cite{yang2021}.
Multiple parameters in noisy quantum circuits can be estimated based on optimal control and reinforcement learning~\cite{xu2020}, avoiding separate optimizations for each parameter.

When the environment of the system changes, the parameters of the Hamiltonian must be recalibrated experimentally as quickly as possible. Different protocols have been suggested to this end. Leveraging concepts from machine learning and optimization, the control parameters of a 53 qubit quantum processor can be calibrated much faster than the system drift~\cite{klimov2020}.
Automated tune-up is possible for any arrangement of coupled qubits~\cite{mortimer2020}. 
Ultimately, for practical device operation, system calibration and control should be unified~\cite{wittler2020}.

\subsubsection{Quantum thermodynamics}
\label{subsec:quthermo}

Control has been an inherent part of thermodynamics. Optimization of the efficiency and power of heat engines has shaped the field. Originally applied to steam engines, \textit{quantum} thermodynamics addresses the issue of miniaturising thermal devices. How small can a quantum heat engine be? What is the optimal performance? Is there a quantum advantage? See~\cite{binder2018,vinjanampathy2016quantum,myers2022quantum-h,pekola2021} for a recent overview. Quantum control and quantum thermodynamics
are closely interlinked:
\begin{enumerate} \renewcommand{\labelenumi}{(\alph{enumi})} 
    \item Thermodynamical consistency restricts the structure of the open system control GKLS master equation~\cite{alicki2018,dann2021non,dann2021quantum,dann2021open,wackerPRA2022}.
    \item Certain control task require a change of entropy, such as reset or thermalization~\cite{ticozziSciRep2014,ticozziQST2017,basilewitschNJP2017,fischerPRA2019,basilewitschPRR2021,dann2019shortcut,dann2020fast}. Tasks that do not require a change of entropy may still benefit from it, for example by reaching the target while actively cooling~\cite{kallush2022,menuPRR2022}.
\item Quantum control  can be used to optimize 
the operation cycle of heat engines and refrigerators~\cite{khait2021,erdman2021identifying,xu2021reinforcement,dann2020fast}.
\item Experimental realizations  of quantum information control and quantum  heat devices share common platforms~\cite{rossnagel2016single,von2019spin,klatzow2019experimental,ono2020analog,guthrie2021cooper,levy2020single}.
\item Quantum thermodynamics supplies a resource theory framework addressing the issue of the cost of the control \cite{Lostaglio19r}.
\end{enumerate}

Optimal control theory requires a dynamical equation of motion connecting the input and the target state. For open quantum systems the theory is based on the GKLS master equation, cf. Sec.~\ref{sec:controllability}.  For realistic quantum devices it is almost impossible to derive from first principle these equations. As a result, an empirical approach for developing quantum simulations prevails, employing the GKLS structure and fitting the Lindblad jump operators and kinetic coefficients to the scenario.
Quantum thermodynamics imposes additional conditions on the admissible dynamical equations.
An underlying assumption is that the dynamics of the system and its surrounding environment is unitary. This assumption is shared with the theory of open quantum systems~\cite{Lind76,Koss72}.
Thermodynamics then imposes additional restrictions on the reduced dynamical  map of the controlled system $\Lambda$. The fixed point of this map should be a thermal equilibrium state with the temperature dictated by the environment. Another idealization is imposed by requiring  an isothermal partition between system and environment.  In this partition no energy is accumulated on the interface. This assumption reflects the intuition that the quantum device is relatively isolated from the environment allowing local measurement of its observables. It is also consistent with a derivation of the master equation based on the weak coupling limit and the secular approximation~\cite{Davies74,scali2021local,trushechkinPRA2021,cattaneo2020symmetry}. In addition, strict energy conservation implies a dynamical time-translation symmetry: The dynamical map of the system commutes with the free unitary map ${\cal U}$, i.e., $[\Lambda, {\cal U}]=0$ \cite{dann2021open,jacob2021thermalization}. Time-translation symmetry implies that the environment  cannot serve as a clock for the system \cite{dann2021non}. From a control prospective, thermodynamics restricts the admissible dynamical equations of motion; the dissipative and unitary parts are linked.
For slow external driving in the adiabatic limit, the free evolution generator composed from the commutator of the instantaneous Hamiltonian and the dissipative generators commute~\cite{alicki2018,albash2012quantum,dann2021open}.
This implies that coherence in the energy frame and population evolve independently.

Rapid control typically described by a time-dependent Hamiltonian requires a non-adiabatic treatment
of the dissipative map~\cite{dann2021quantum}. Time translation symmetry imposes the condition that the free unitary map and the dissipative map commute. A procedure to obtain the generators of the GKLS master equation has been developed based on the inertial theorem \cite{dann2021inertial}. To date, this procedure has been obtained only for closed form solution of the free dynamics~\cite{dann2019shortcut,dann2020fast}.

Controlling the Hamiltonian directly influences the unitary evolution  accompanied by an indirect control of the dissipation. This dependency influences the rules for open system controllability presented in Sec.~\ref{sec:controllability}.
Under these conditions, it can be inferred that systems that are unitary controllable are state to state controllable~\cite{dann2020fast,dirr2019}. An open problem are the controllability criteria for dynamical maps $\Lambda_f$ under thermodynamically consistent dissipation.

Entropy-changing transformations are a hallmark of quantum thermodynamical control tasks. An elementary and universal task is the reset transformation. The control objective is a fast reset to a desired state with high fidelity~\cite{ticozziSciRep2014,ticozziQST2017,basilewitschNJP2017,fischerPRA2019,basilewitschPRR2021}. The fidelity is restricted by the third law of thermodynamics: Very high fidelity requires infinite resources~\cite{taranto2021landauer}.
The speed of the reset dynamical map is related to the rate of transfer of entropy to the environment.
Reset and cooling are similar tasks for optimization of entropy-changing transformations. Cooling via a delta kick protocol has been proposed~\cite{dupaysPRR2021} where the control is achieved by switching on and off a noise source. These reset and cooling mechanisms are in line with the reachable set when the interaction with the environment is controllable, cf. subsection \ref{sec:Reach-Markov-Open}.
A related control task is to speed up equilibration. A control strategy based on the inertial theorem was developed~\cite{dann2019shortcut,dann2020fast}. 
Speed-up is obtained by maintaining the controlled system as far from equilibrium as possible by generating significant coherence. The final step is to rotate this coherence back to population. The speedup comes with a cost, extra work is required which is dissipated to the environment producing entropy~\cite{dann2020quantum}.

Preserving entropy or, minimizing decoherence is another legitimate control task.  When considering protecting quantum gates from dissipation, active cooling is possible while performing the quantum operation~\cite{kallush2022}. 
An alternative strategy is to formulate the control  in a decoherence free subspace~\cite{mousolou2018realization,hu2021optimizing,wu2021nonadiabatic},
or in a path independent control
\cite{ma2020path,ma2022algebraic}.

A favored subject of QOCT in quantum thermodynamics are heat engines---devices that convert heat to work or operate in reverse as refrigerators \cite{myers2022quantum-h}. These engines can be classified as autonomous, continuously driven, and discrete~\cite{binder2018}. 
QOCT for heat engines has been almost exclusively applied to the discrete Carnot and Otto four stroke cycles. Typical optimization targets are maximum efficiency, maximum power, or minimal fluctuations~\cite{van2021thermodynamics,ye2022optimal,erdman2022}. A trade-off has been identified between these tasks. Optimizing power requires reducing the engine's cycle period. Typically, this protocol is accompanied by an increase in dissipation and therefore
reduced efficiency. 

In small quantum engines fluctuations become important~\cite{hasegawa2021thermodynamic,miller2021thermodynamic,koyuk2021quality,liu2021coherences,liu2021coherences,boutonNATURE2021}, diverging fluctuations make the device useless. Actively controlling fluctuations comes at the expense of either efficiency or power.
In view of miniaturizing quantum devices the issue of fluctuation will become more important. Active control
to reduce fluctuations will become a legitimate goal
\cite{shastri2022}.

The Otto cycle has been a popular target for optimization. The cycle is composed of two unitary branches and two thermalization branches, thus separating the controlled segment from the dissipation.
The unitary branches are characterized by rescaling  the Hamiltonian.  Whenever the drift Hamiltonian does not commute with the control operators rapid  protocols will generate coherence.
Generating coherence from an initial thermal state has a cost in additional work~\cite{kallush2019quantifying}. If coherence is present at the terminus of the unitary stroke, the extra work will be dissipated in the thermalization strokes reducing the engine's efficiency. This phenomena is termed quantum friction~\cite{feldmann2003quantum,kosloff2010optimal,reiche2020nonequilibrium,insinga2020quantum,ccakmak2020quantum,miller2021thermodynamic}. The friction loss has been the motivation for optimizing the protocol for the unitary branches~\cite{salamon2009maximum,insinga2020quantum,stefanatos2021shortcut,singh2021unified}. Some protocols employ shortcuts to adiabaticity, cf. Sec.~\ref{subsec:shortcuts}, since at the terminus of the stroke no coherence is present~\cite{del2014more,alipour2020shortcuts,dupaysPRR2021,abahPRR2020}. Examining these protocols, coherence is generated but is transient. There is a dispute if to associate a cost to this coherence~\cite{abahNJP2019,torronteguiPRA2017,kiely2022}. To overcome this cost, control methods were applied, for example a combination of dynamic programming, machine learning and STA~\cite{khait2021,erdman2021identifying}.

Optimizing the  Carnot cycle requires  control of the isothermal strokes of the engine~\cite{dann2020quantum,dann2020quantum2}. In this stroke, the  Hamiltonian is varying  while the working medium is in contact with the thermal bath. The introduction of the non-adiabatic Master equation~\cite{dann2018time} enabled to
study the cycle and its optimization. The optimized quantum Carnot cycle was found to possess the typical trade-off between
power and efficiency \cite{dann2020quantum}. Speeding up the thermalization resulted in an increase in dissipation. These studies were based on the inertial approach allowing a quasi-analytic solution~\cite{dann2021inertial}. In the weak dissipation limit for slow driving, a geometric optimization approach was employed for a general engine cycle~\cite{abiuso2020geometric,ma2021consistency}. 
The basic idea is to minimize a distance metric of the cycle from the equilibrium state. This approach has also been employed to  minimize dissipation
\cite{deffner2020thermodynamic}.

Experimental realizations of quantum heat engines employ the typical setups developed for quantum information processing, such as ion traps~\cite{rossnagel2016single,von2019spin,hu2020quantum}, 
cold atoms \cite{boutonNATURE2021}, NV centers~\cite{klatzow2019experimental}, and Josephson devices~\cite{ono2020analog,guthrie2021cooper,campisi2020}, 
including a realization of an absorption refrigerator~\cite{gubaydullin2021photonic}.
These studies demonstrate that quantum thermodynamic principles are able to address devices ranging from macroscopic bulk engines to single qubit operated devices.

In quantum devices accounting for the physical resources required 
to achieve the control task is a fundamental issue. Thermodynamics as a theory has been
constructed to address this issue. Quantum thermodynamics is a confluence of quantum information, quantum statistical mechanics and quantum dynamics. The theory
therefore has the capability to assess the requirements for quantum control.
In optimal control theory 
the energetic cost is typically incorporated by employing a Lagrange parameter. 
The invested energy in the control is closely associated with the quantum speed limit, cf. Sec.~\ref{sec:Q-Speed-Limit}. 
The  energetic cost in quantum thermodynamics is reflected by the first law  \cite{sparaciari2020first}; and the cost of quantum gates has been analyzed~\cite{deffner2021energetic,aifer2022quantum}. A different viewpoint is accounting for irreversible entropy generation required for control \cite{kiely2022,kallush2022}. 
The real resource for control is coherence.
Under a unitary map coherence is preserved. 
If we incorporate the system and controller in a super-quantum system, coherence is transferred from the controller to the system. In the semiclassical limit  the controller is described by a time-dependent field. 
This framework supports the viewpoint that coherence is a resource~\cite{streltsov2017colloquium}.
A quantum signature in heat engines is the conversion of coherence to useful work~\cite{uzdin2015equivalence,francica2020quantum}.

\subsection{Goals and challenges for advancing the application of QOCT}

The impressive progress in the application of QOCT to the various hardware platforms and control tasks paves the way to further extending the versatility of the QOCT toolbox. This requires, at the same time, significant advances from control hardware all the way to new conceptual solutions. A key challenge is hardware development of a scalable control architecture; a microarchitecture for efficient instruction-driven pulse synthesis has just been brought forward~\cite{khammassi2022}.
Another key challenge is a better integration of control and calibration. For example, a requirement on future control electronics is a powerful internal optimization logic that allows for fast pulse calibration. At the same time, better quantum engineering in the sense of isolating and protecting quantum systems from external noise will continue to be an active field for control and optimization.

Further progress is also required in basic control tasks, in particular those that cannot be achieved with purely coherent control. For example, a prerequisite for a quantum device is cooling or reset to a purified initial state. Control and optimization of these processes carries a substantial benefit and this subject will continue to be a major research topic in the near future. Similarly, a quantum refrigerator removing entropy from the sensing or computation part supplying cold ancillas is likely part of future technology~\cite{clivaz2019unifying,taranto2020exponential,gluza2021quantum,solfanelli2022}.

At the more conceptual level, QOCT in open quantum systems has so far largely been based on a Markovian framework to supply the dynamical equations of motion, cf. Sec.~\ref{sec:Markovianity}.  An open problem is a thermodynamically consistent theory for non-Markovian, driven dynamical systems~\cite{dann2021non} that can be combined with QOCT, i.e., with arbitrarily fast drives.

\section{Long-term vision for quantum optimal control in quantum technologies}\label{sec:vision}

The vision for the future of quantum technologies is for quantum devices to provide advantages with respect to classical devices in a broad range of applications.
In this context, the mission of quantum optimal control methods is to make it possible for quantum technological devices to reach and maintain their best performance outside of the lab, under real-world conditions.

This concerns the operation of quantum devices at many levels. 
Resource management is a crucial element in quantum engineering.
In the case of quantum computing, for instance, the aims of future QOCT encompass almost the whole software stack underneath high-level programming languages. This ranges from expanding the library of variational and optimisation-based algorithms such as QAOA and VQE to orchestrating the distribution of computational tasks between classical and quantum co-processors, and from compiling quantum circuits for reduced complexity of the required gate sequences to enhancing the effectiveness of standard quantum control tasks such as pulse shaping for hardware optimisation. 

Several of these goals are essential also for the deployment of protocols for quantum communication and quantum sensing. To fulfil the needs and realise the potential of all QT application areas, future QOCT will aim at providing calibration of the control sequences to the specific parameters of individual devices, as well as recalibration for adapting to parameter drifts and other systematic disturbances, in a fully automated manner not requiring constant specialist intervention. 

To this end, QOCT aims not only at building on efficient and reliable modelling and system identification, but also at consistently improving the models it relies on based on the gathered data. This will require to intensify the already increasingly pervasive use of machine learning techniques. The final goal will be the achievement of general-purpose, specifically adaptable tools --- a universal toolbox ideally to be  automatically tailored to the particular physical configuration of any given quantum device to enable attainment of the best possible performance. Along the process, this will include the ability to prescribe the most suitable requirements for the underlying quantum hardware, leading to a systematisation of co-design, and ultimately to a true quantum systems theory.

Finally, another key aspect in the development of quantum technologies through the large-scale application of quantum control techniques will be the training of engineers, researchers, and students in this rapidly evolving field both from the experimental and theoretical points of view. Many initiatives are currently proposed, in Europe and elsewhere, to improve the quantum workforce education. For the quantum control part of that education, our roadmap provides an overview from which essential components of the future common knowledge framework can be drawn. 

Novel challenges will keep the field alive and vibrant. At the same time, it is an important task for the quantum control community to make its tools accessible to a wide audience with different levels of technical skills. This is key to ensuring the long-term impact of quantum optimal control, which is to become the underlying basis for any quantum technology application naturally embedded into each and every quantum device. 

\section{Conclusions}\label{sec:conclusions}

We have reviewed the current state of the art in quantum optimal control as relevant to the fast evolving field of quantum technologies and summarized the most pressing open questions, updating an earlier roadmap~\cite{glaserEPJD2015}.
When inspecting the global perspective that our overview provides, two observations are striking:

(i) QOCT has significantly matured over the past few years and is about to become a routine tool for laboratory quantum technologies. The next step will be to push this development towards even more versatility and user-friendliness, to allow for integration in practical quantum devices at the application-ready level. The need for this development has already been realized, and more traditional academic research settings are now being complemented by industrial development. 

(ii) There has been a lot of cross-fertilization with neighbouring fields, with machine learning in both their classical and quantum versions as prominent example, but there is plenty of room for more. As the quantum technologies spread out towards engineering and computer science, there is, at the same time, a further need to unify the various languages, or rather dialects, that capture the very same foundational concepts. Take the example of the figure of merit --- it is the target functional in QOCT, the fidelity in system and process characterization, and the cost in machine learning and the question about the resources needed for their estimation. A global overview like the one we are presenting here will hopefully serve to identify commonalities and thus prepare the ground for further cross-fertilization.

Such cross-fertilization suggests that the long-term future of quantum optimal control is to be an integral part of the larger technical foundations of the quantum technologies.


\begin{backmatter}

\section*{Acknowledgements}

We would like to thank  D. Gu\'ery-Odelin for useful comments and Josias Langbehn for technical help. 

\section*{Funding}
Financial support from European Union's Horizon 2020 research and innovation programme under the Marie Sk{\l}odowska-Curie grant agreement Nr. 765267 (QuSCo) is gratefully acknowledged. 
SF, SJG and TSH would also like to acknowledge the Munich Quantum Valley (MQV) e.V., München, Germany.

\section*{Abbreviations}

Arbitrary Wave form Generator (AWG);
Bose-Einstein Condensate (BEC);
Chopped RAndom Basis (CRAB);
Completely Positive (CP);
Completely Positive and Trace-Preserving (CPTP);
Derivative Removal by Adiabatic Gates (DRAG);
Greenberger, Horne, Zeilinger (GHZ);
Gorini, Kossakowski, Lindblad, and Sudarshan (GKLS);
GRadient Ascent Pulse Engineering (GRAPE);
Gradient Optimization of Analytic conTrols (GOAT):
Electron Paramagnetic Resonance (EPR); 
Nitrogen Vacancy (NV) Centers;
Noisy Intermediate-Scale Quantum (NISQ);
Nuclear Magnetic Resonance (NMR);
Pontryagin Maximum Principle (PMP);
Quantum Approximate Optimization Algorithm (QAOA);
Quantum ElectroDynamics (QED);
Quantum Optimal Control Theory (QOCT);
Quantum Speed Limit (QSL);
Quantum Technologies (QT);
Reinforcement Learning (RL);
Shortcuts To Adiabaticity (STA);
STImulated Raman Adiabatic Passage (STIRAP);
Variational Quantum Eigensolver (VQE);



\section*{Competing interests}
The authors declare that they have no competing interests.


\section*{Authors' contributions}
The compilation of this review was coordinated by Christiane Koch. 
Section~\ref{sec:controllability} was written by Ugo Boscain, Gunther Dirr, and  Thomas Schulte-Herbr\"uggen; Sections~\ref{sec:methods} and ~\ref{sec:comparison} by Steffen Glaser, Christiane Koch, Dominique Sugny, and Frank Wilhelm; and Section~\ref{sec:applications} by Stefan Filipp, Steffen J. Glaser, Christiane Koch, Ronnie Kosloff, Simone Montangero, Dominique Sugny, and Frank Wilhelm. All authors contributed to the writing of Sections~\ref{sec:intro}, \ref{sec:vision} and \ref{sec:conclusions} and to the revision of the complete manuscript.



\bibliographystyle{abbrvnat} 
\setcitestyle{number,open={[},close={]}}
\bibliography{CtrlRdmpUpdate,control21short}      

\end{backmatter}
\end{document}